\documentclass[fleqn,10pt]{wlscirep}
\usepackage[utf8]{inputenc}
\usepackage[T1]{fontenc}
\usepackage{siunitx}
\newcolumntype{M}[1]{>{\raggedright\arraybackslash}p{#1cm}}

\usepackage{pifont}
\newcommand{\cmark}{\ding{51}}%
\newcommand{\xmark}{\ding{55}}%

\usepackage{array}
\newcolumntype{P}[1]{>{\centering\arraybackslash}p{#1}}

\title{QMugs: Quantum Mechanical Properties of Drug-like Molecules}

\author[1,$\dag$]{Clemens Isert}
\author[1,$\dag$]{Kenneth Atz}
\author[1,2,*]{José Jiménez-Luna}
\author[1,3,*]{Gisbert Schneider}
\affil[1]{Department of Chemistry and Applied Biosciences, RETHINK, ETH Zurich, 8093 Zurich, Switzerland.}
\affil[2]{Department of Medicinal Chemistry, Boehringer Ingelheim Pharma GmbH \& Co. KG, Birkendorfer Straße 65, 88397 Biberach an der Riss, Germany.}
\affil[3]{ETH Singapore SEC Ltd, 1 CREATE Way, $\#$06-01 CREATE Tower, Singapore 138602, Singapore.}

\affil[*]{corresponding authors: Gisbert Schneider (gisbert@ethz.ch), José Jiménez-Luna (jose.jimenez@rethink.ethz.ch)}
\affil[$\dag$]{these authors contributed equally to this work}

\begin{abstract}

Machine learning approaches in drug discovery, as well as in other areas of the chemical sciences, benefit from curated datasets of physical molecular properties. However, there is a lack of sufficiently large data collections that include first-principle quantum chemical information on bioactive molecules, such as single-point electronic properties, quantum mechanical wave functions and density-functional theory (DFT) matrices. The open-access QMugs (Quantum-Mechanical Properties of Drug-like Molecules) dataset fills this void. The QMugs collection comprises quantum mechanical properties of more than 665k biologically and pharmacologically relevant molecules extracted from the ChEMBL database, totaling $\sim$2M conformers. QMugs contains optimized molecular geometries and thermodynamic data obtained via the semi-empirical method GFN2-xTB. Atomic and molecular properties (\textit{e.g.}, partial charges, energies, and rotational constants) are provided on both the GFN2-xTB and on the DFT ($\omega$B97X-D/def2-SVP) levels of theory. QMugs also comprises the respective quantum mechanical wave functions, including DFT density and orbital matrices, totaling over 7 terabytes of uncompressed data. This dataset is intended to facilitate the development of models that learn from molecular data on different levels of theory while also providing insight into the corresponding relationships between molecular structure and biological activity.

\end{abstract}
\begin{document}

\flushbottom
\maketitle
\thispagestyle{empty}
\section*{Background \& Summary}

Machine learning methodologies are increasingly becoming well-established tools in many chemistry-related disciplines, such as drug discovery~\cite{gawehn2016deep}, material science~\cite{schmidt2019recent}, and physical chemistry~\cite{von2018quantum}. In recent years, significant progress has been made in quantum-based machine learning (QML) methods~\cite{von2020exploring}, which aim to accurately and computationally inexpensively predict the governing properties of atomistic systems, such as energies and forces~\cite{satorras2021n, schutt2021equivariant, huang2020quantum, christensen2020fchl, heinen2020quantum, heinen2020machine, christensen2019operators, faber2018alchemical}, dipole moments~\cite{balcells2020tmqm}, wave functions~\cite{unke2021se3equivariant, schutt2019unifying} and electron densities~\cite{grisafi2018transferable, fabrizio2019electron}. Despite the success and promise surrounding the applicability of such approaches, several challenges remain for QML. Arguably, one of the most important challenges is the increasing need for curated, comprehensive datasets.~\cite{balcells2020tmqm} While several options, such as the QM9~\cite{ramakrishnan2014quantum} or ANI-1~\cite{smith2017anid} sets have paved the way for the development of current-generation QML methods~\cite{satorras2021n, schutt2021equivariant, huang2020quantum, christensen2020fchl, smith2017ani, smith2020ani, nakata2020pubchemqc}, the computational cost entailed in their generation limits both the scope of the explored chemical space (\textit{e.g.}, molecule size, atom-type diversity), and prospective modeling applicability~\cite{balcells2020tmqm, glavatskikh2019dataset}. 

There has been a recent surge in interest in the delta-learning ($\Delta$-learning) of chemical properties, which aims to use a machine learning model to predict a physically relevant quantity, such as those generated by density-functional theory (DFT) by utilizing information extracted with a computationally cheaper method~\cite{qiao2020orbnet, smith2020ani} (\textit{e.g.}, semi-empirical approaches such as GFN2-xTB~\cite{grimme2017robust, bannwarth2019gfn2, pracht2019robust, grimme2019exploration, bannwarth2020extended} and PM6~\cite{rezac2009semiempirical}). Datasets that enable this type of learning are scarce and could promote the development of accurate models at potentially a fraction of the computational cost of more precise alternatives~\cite{folmsbee2021assessing}. Furthermore, datasets that provide three-dimensional conformational data, for a wide variety of chemical space, at levels of theory higher than classical force fields~\cite{bolton2011pubchem3d, axelrod2020geom}, could boost the performance of machine learning methods in predicting properties from ensembles as well as generative models of conformations. Relevant examples include the PubChemQC-PM6~\cite{nakata2020pubchemqc} and GEOM~\cite{axelrod2020geom} datasets, which include molecules with properties computed using different semi-empirical levels of theory. Finally, there is a clear potential to open up new lines of research by combining biological annotations (\textit{e.g.}, from molecular databases such as ChEMBL~\cite{mendez2019chembl}), and additional QM-derived physical information.

\begin{figure}[ht]
\centering
\includegraphics[width=\linewidth]{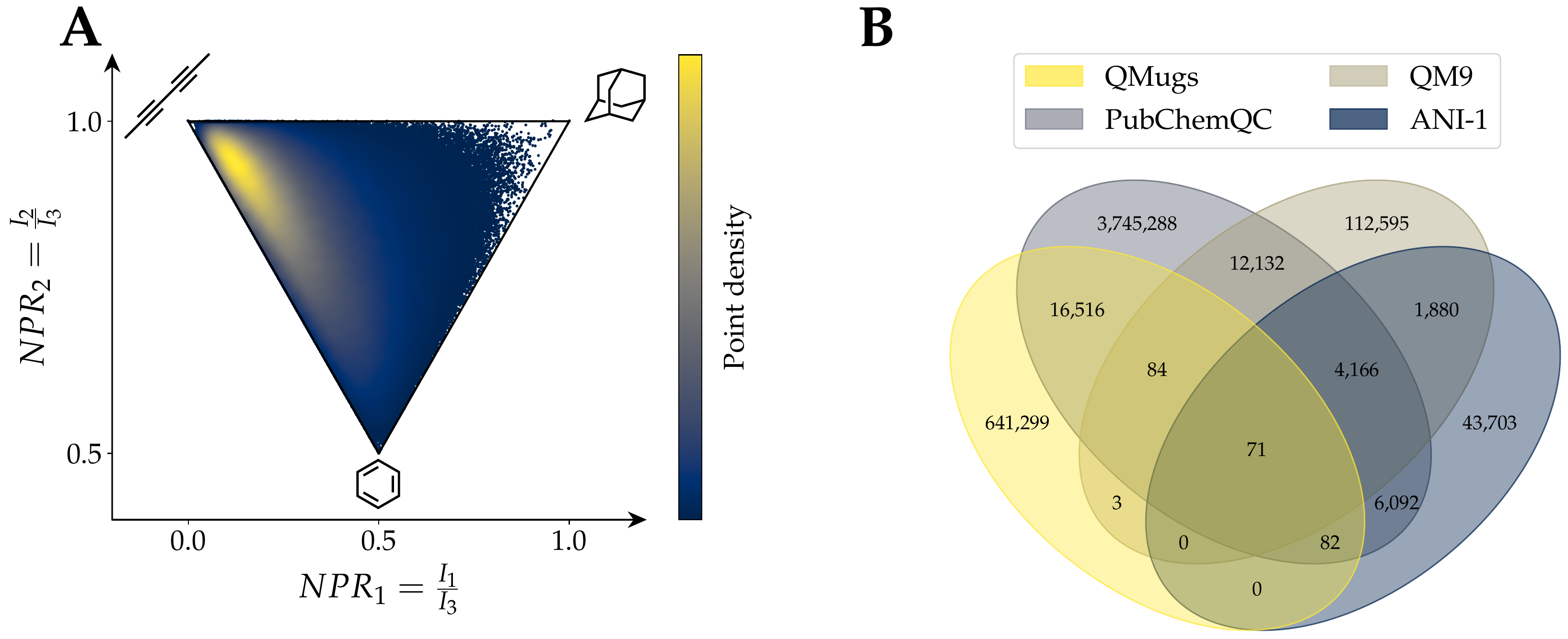}
\caption{\textbf{A})~Principal-moments-of-inertia plot~\cite{sauer2003molecular} for molecules in the QMugs dataset. $NPR_x$ = $x$-th normalized principal moment, $I_x$ = $x$-th smallest principal moment of inertia.
\textbf{B})~Venn diagram showing overlap between QMugs and other well-known datasets with DFT-level computed properties: QM9~\cite{ramakrishnan2014quantum}, PubChemQC~\cite{nakata2017pubchemqc}, and ANI-1~\cite{smith2017anid}. Overlap was computed based on the uniqueness of the InChI representations of the contained molecules. Numbers do not add up to those reported in Table~\ref{tbl:desc} because of InChI strings that occur multiple times.}
\label{fig:rod_disk_sphere_venn}
\end{figure}

\begin{figure}[ht]
\centering
\includegraphics[width=\linewidth]{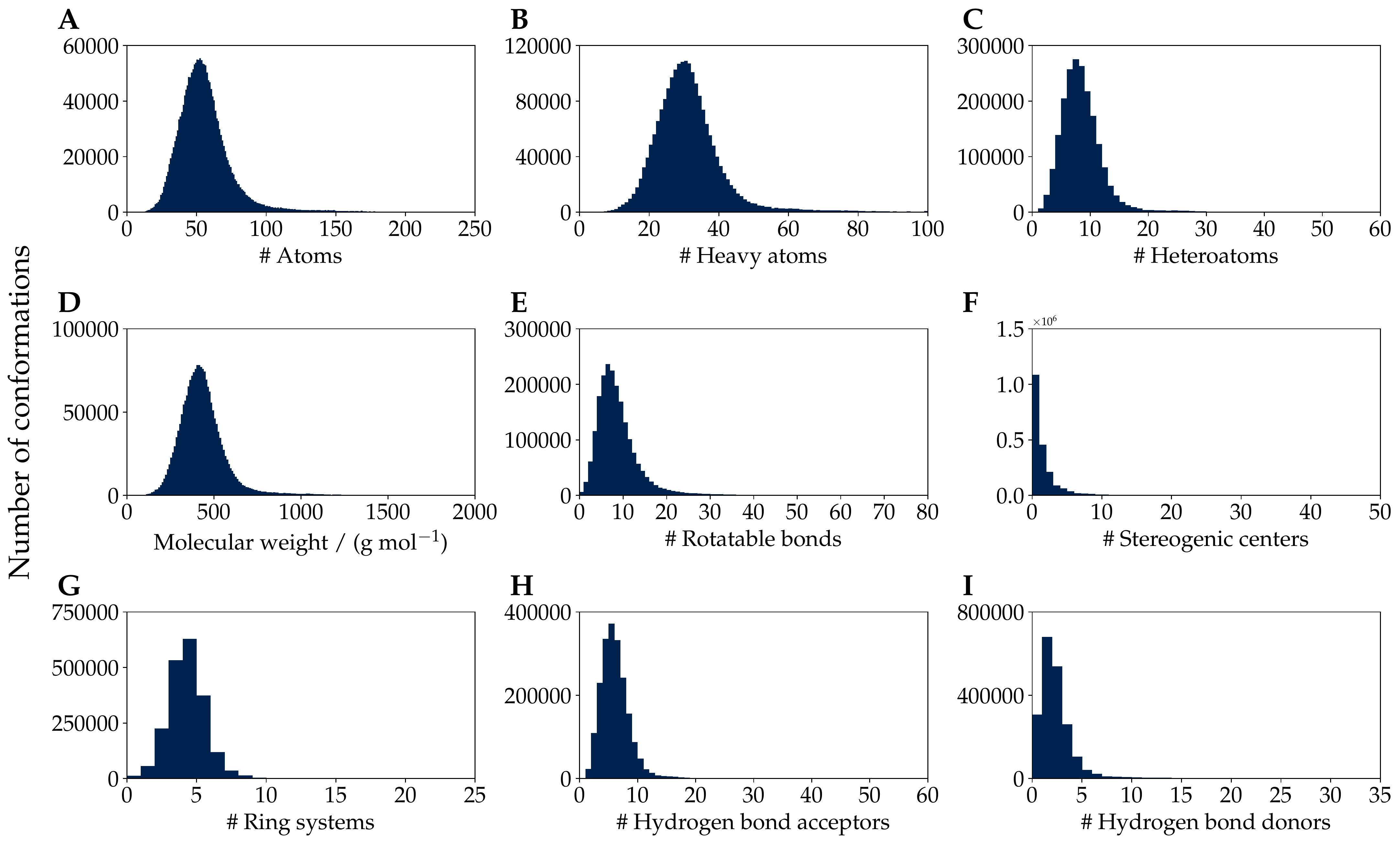}
\caption{Distribution of properties for the molecules contained in the QMugs dataset.}
\label{fig:molecule_props}
\end{figure}

This work introduces QMugs (\textbf{Q}uantum-\textbf{M}echanical Properties of Dr\textbf{ug}-like Molecule\textbf{s}), a data collection of over $665$k curated molecular structures extracted from the ChEMBL database, with accompanying computed quantum mechanical properties. Different levels of theory were combined in these calculations. Per compound, three conformers were generated, and their geometries were optimized using the semi-empirical GFN2-xTB method~\cite{grimme2017robust, bannwarth2019gfn2, pracht2019robust, grimme2019exploration, bannwarth2020extended}, whereas a comprehensive array of quantum properties was computed at the DFT level of theory using the $\omega$B97X-D functional~\cite{chai2008long} and the def2-SVP Karlsruhe basis set~\cite{weigend2005balanced}. The data collection presented herein is put in the context of other sets that also feature DFT-level properties. A descriptive evaluation against QM9~\cite{ramakrishnan2014quantum}, PubchemQC~\cite{nakata2017pubchemqc}, and the ANI-1~\cite{smith2017anid} datasets is provided in Figure~\ref{fig:rod_disk_sphere_venn}B, as well as in Table~\ref{tbl:desc}. As previously reported for the ChEMBL database~\cite{meyers2016origins}, most of the considered drug-like molecules in this study fall within the rod-disk axis in the principal moments of inertia plot~\cite{sauer2003molecular} (Figure~\ref{fig:rod_disk_sphere_venn}A). Furthermore, the vast majority of the included compounds ($\sim 641$k, $96.3\%$) were previously unreported in other DFT data collections, while also providing equivalent information at additional levels of theory, namely GFN2-xTB. With an average of $30.6$ and a maximum of $100$ heavy atoms per compound (Table~\ref{tbl:desc} \& Figure~\ref{fig:molecule_props}), QMugs also features molecular samples that are considerably larger than those provided by other datasets. To the best of our knowledge, this work is the first to provide a large and diverse dataset of quantum mechanical wave functions represented as local bases of atomic orbitals (\textit{i.e.}, DFT density and orbital matrices). Single-point properties as well as wave functions were computed with the Psi4 software suite~\cite{smith2020psi4} for all the conformers ($\sim 2.0$M) present in the database, totaling over $7$ terabytes of supplied quantum mechanical data.

Overall, the utility of the presented dataset is fourfold: (i) it will provide researchers with the largest-to-date dataset to either directly predict the quantum chemical properties, or learn a property mapping between two popular quantum mechanical levels of theory (\textit{i.e.}, GFN2-xTB and $\omega$B97X-D/def2-SVP); (ii) it will facilitate the development of novel machine-learning methodologies for the generation of molecular conformations and molecular property predictions via their ensembles; (iii) it will enable the development of novel deep learning frameworks for the prediction of the quantum mechanical wave function in a local basis of atomic orbitals from which all ground-state properties as well as electron densities can be derived; and (iv) it will enable research towards the exploration of QML methods quantum featurization in the context of pharmacologically relevant, annotated biological data.

\section*{Methods}
\begin{figure}[ht]
\centering
\includegraphics[width=\linewidth]{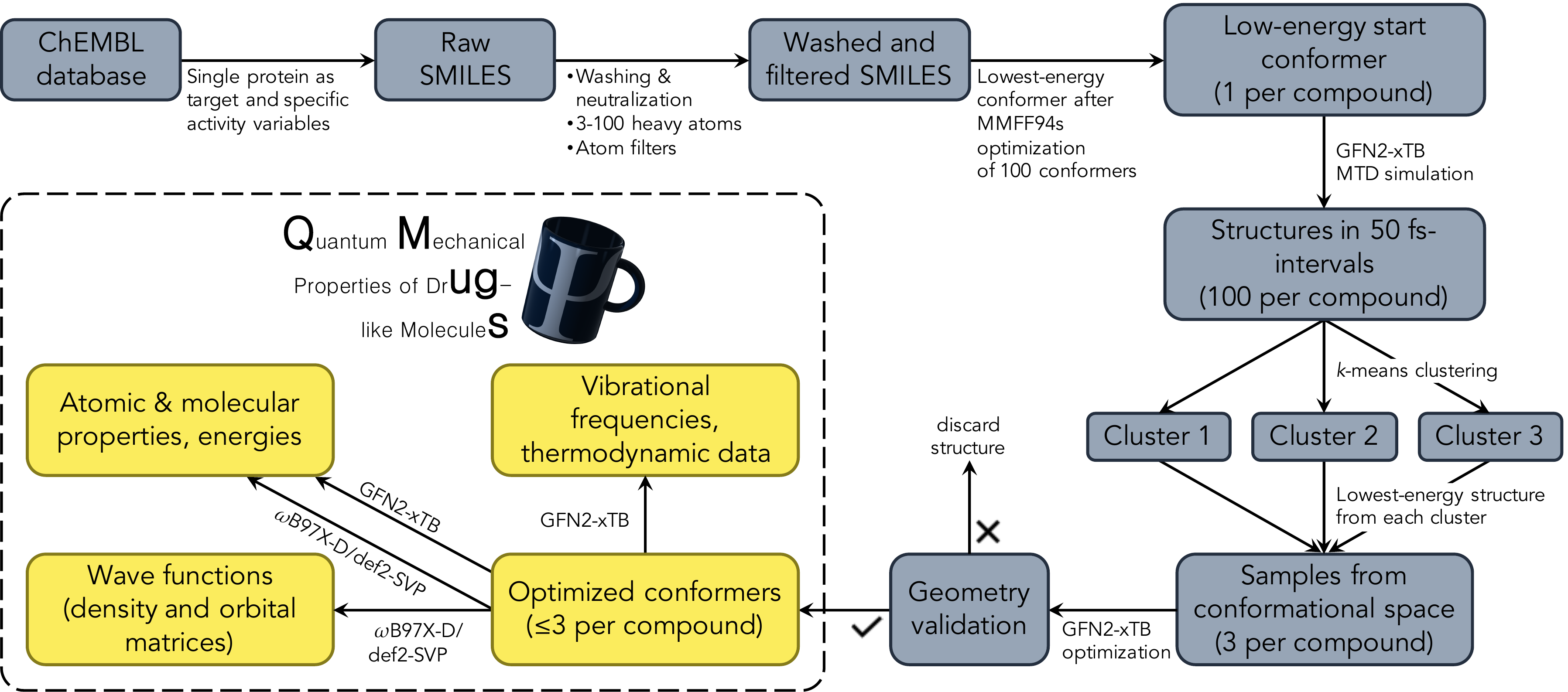}
\caption{Overview of the data generation process. Molecules were extracted from the ChEMBL database, standardized, and filtered, and starting conformers were generated using the RDKit software package. Metadynamics (MTD) simulations were performed using the GFN2-xTB semi-empirical method to generate three diverse conformations before final geometry optimization. Molecules that did not pass a series of geometric sanity checks were removed. DFT-level properties ($\omega$B97X-D/def2-SVP) were computed using the Psi4 software.}
\label{fig:pipeline}
\end{figure}

Molecules were extracted from the ChEMBL database~\cite{mendez2019chembl} (version 27). Conformers were generated using RDKit~\cite{rdkit} and GFN2-xTB~\cite{grimme2017robust, bannwarth2019gfn2, grimme2019exploration, bannwarth2020extended}. DFT ($\omega$B97X-D/def2-SVP) calculations were carried out via Psi4~\cite{smith2020psi4}. A similar approach was adopted in a previous study on transition-metal complexes.~\cite{balcells2020tmqm} An overview of the data processing pipeline is given in Figure~\ref{fig:pipeline}, while individual steps are described in more detail in the following subsections.

In chemical terminology, the term ``conformation'' refers to any arrangement of atoms in space, whereas ``conformer'' refers to a conformation that is a local minimum on the potential energy surface of the molecule.~\cite{moss1996basic} In the analyses that follow, the term ``conformation'' is loosely used to refer to both, unless explicitly mentioned otherwise.

\subsection*{Data extraction and SMILES processing}

Single-protein targets with assay information for at least $10$ compounds with unique internal identifiers were extracted from the ChEMBL database. Several activity and annotation filters were subsequently applied to these compounds (see ESI for a detailed query description). This procedure resulted in $685,917$ molecules with unique external identifiers (ChEMBL-IDs), represented by their Simplified Molecular Input Line Entry Specification (SMILES)~\cite{weininger1988smiles}. Molecules were neutralized, and salts and solvents were removed using the ChEMBL Structure Pipeline package~\cite{Bento2020, chembl_structure_pipeline_repo}. For compounds consisting of multiple separate fragments after this "washing" procedure, all except the one with the highest number of heavy atoms were discarded. Additionally, molecules containing fewer than $3$ or more than $100$ heavy atoms, as well as radical species and molecules with a net charge different from zero after the attempted neutralization, were removed. Atom types included in the QMugs dataset are hydrogen, carbon, nitrogen, oxygen, fluorine, phosphorus, sulfur, chloride, bromide, and iodine.

\subsection*{Conformer generation and optimization}

With the procedure described herein, a compromise between efficient molecular conformational search and practical computational expense considerations was sought. 

The RDKit~\cite{rdkit} implementation of the Experimental-Torsion Knowledge Distance Geometry (ETKDG) method~\cite{riniker2015better} was used to generate up to $100$ conformers for each molecule, with a maximum of $1000$ embedding attempts and an initial coordinate assignment using distance-matrix eigenvalues and default settings (\texttt{boxSizeMult=2.0}, \texttt{force-field tolerance=1e-3}) . Upon no successful conformer generation, it was re-attempted via random assignment of the starting coordinates. The resulting conformers were further minimized using the Merck molecular force field~\cite{tosco2014bringing} (MMFF94s) for a maximum of $1000$ iterations, with default settings (\texttt{nonBondedThresh=100.0}). The lowest-energy conformer (according to the selected force field) for each structure was then used as a starting point for meta-dynamics (MTD) simulations. Stereocenters that were previously undefined in the SMILES extraction procedure were assigned in this conformer generation process.

For each generated conformer, an MTD simulation was performed with the xTB software package~\cite{grimme2019exploration} for a duration of $5$ ps with time steps of $1$ fs, at a temperature of $300$ K. The biasing root-mean-square deviation (RMSD) potential used for all MTD simulations is given by $E_{\mathrm{bias}}^{\mathrm{RMSD}} = \sum_{i=1}^{N} k_{i} \exp(\alpha \Delta^{2}_{i}$), where $N$ is the number of reference structures, $k_i$ the pushing strength, $\Delta_i$ the collective variable (\textit{i.e.}, the RMSD between structure $i$ and a reference structure), and $\alpha$ the width of the Gaussian potential used in the RMSD criterion. Simulations were carried out with $\alpha=$\SI{1.2}{\bohr}$^{-1}$ and $k_{i}=$\SI{0.2}{\milli\hartree} with snapshots taken every $50$ fs, resulting in $100$ conformations stored with their corresponding energies. To obtain conformationally diverse samples, these structures were subsequently clustered into three groups via the $k$-means \cite{lloyd1982least} algorithm, as implemented in the scikit-learn~\cite{scikit-learn} (version 0.23.1) Python package using the pairwise RMSD of the aligned structures as molecular features. The conformation with the lowest-energy value from each cluster was then selected for further processing. The three resulting conformations for each molecule were then optimized using the GFN2-xTB~\cite{grimme2017robust, bannwarth2019gfn2, grimme2019exploration, bannwarth2020extended} method using energy and gradient convergence criteria of $5\times 10^{-6}$ \si{\hartree} and $1\times 10^{-3}$ \si{\hartree} $\alpha^{-1}$, respectively, and the approximate normal coordinate rational function optimizer (ANCopt). Harmonic frequencies, entropies, enthalpies and heat capacities at $298.15$ K were extracted at the end of the geometry optimization process. Structures for which vibrational frequencies with imaginary wave numbers were obtained --- indicative of failure to reach energy minima --- were subjected to additional optimizations until no significant ones remained, up to a maximum of $100$ attempts.

\subsection*{Quantum mechanical calculations}

Single-point electronic calculations were performed for the optimized geometries using the $\omega$B97X-D quantum functional and the def2-SVP basis set as implemented in the open-source quantum-chemistry software suite Psi4~\cite{smith2020psi4}. Single-point properties such as formation and orbital energies, dipole moments, rotational constants, partial charges, bond orders, valence numbers, as well as wave functions including $\alpha$ and $\beta$ DFT-density matrices, orbital matrices, and the atomic-orbital-to-symmetry-orbital transformer matrix were obtained. For practical reasons, $52$ structures whose DFT calculations required computational resources that exceeded empirically determined limits, or for which calculations were unsuccessful, were discarded (see ESI for details). 

\section*{Data Records}

All computed molecular structures, as well as their corresponding properties and wave functions are accessible through the ETH Library Collection service (\href{https://doi.org/10.3929/ethz-b-000482129}{Data Citation 1})~\cite{qmugs_repository}.

\subsection*{Format specification}
A \texttt{summary.csv} comma-separated file contains computed molecular-level properties and additional annotations. A compressed tarball file (\texttt{structures.tar.gz}) of $\sim7$~gigabytes (GB) contains plain MDL structure-data files~\cite{dalby1992description} (SDFs) with embedded atomic and molecular properties, grouped in sub-directories according to their respective ChEMBL identifiers. These SDFs include single-point electronic properties calculated on the GFN2-xTB and $\omega$B97X-D/def-SVP levels of theory, as described in Table~\ref{tbl:properties}. A second compressed tarball file (\texttt{vibspectra.tar.gz}, $\sim3$~GB) contains vibrational spectra.

Wave function files, including additional properties such as DFT-density and orbital matrices, as described in Table~\ref{tbl:wfn}, are split into $100$ compressed tarballs (\texttt{wfns\_xx.tar.gz}) of $\sim50$~GB each for easier management and downloading. These are supplied as NumPy~\cite{harris2020array} (\texttt{.npy}) binary files, which can be read using the Psi4 software package. Molecules (with all conformers grouped together) were assigned at random to the tarballs to enable easy use of subsets of the QMugs dataset without having to download all the files. The assignment of ChEMBL identifiers to tarballs is described in a \texttt{tarball\_assignment.csv} file.

\section*{Technical Validation}

\subsection*{Optimized geometry sanity checks}
Four consecutive geometry checks were performed to filter out structures for which the geometry optimization procedure converged to unrealistic conformations. To determine suitable thresholds for removing a structure from our dataset, the generated geometries were compared to experimental reference values and to DFT-optimized geometries extracted from the PubChemQC dataset~\cite{nakata2017pubchemqc}. Specifically, we investigated (i) the deviation of bond lengths from experimental reference values, (ii) isomorphism between the initial molecular graphs and those obtained after geometry optimization, (iii) linearity of triple bonds, and (iv) planarity of aromatic rings. Structures were removed from the dataset if they failed any test of these tests. In total, $10,986$ ($0.55$\%) conformations were discarded from the dataset. Each test is briefly described in the following subsections, with further technical details reported in the ESI.

\subsubsection*{Deviation of bond lengths from experimental reference values}
Bond lengths in the optimized structures were compared to average experimental reference values for bonds of the same bond type (single, double, triple, or aromatic) and between the same atoms. Reference values were obtained from the Computational Chemistry Comparison and Benchmark DataBase (CCCBDB)~\cite{nist_database}, and the largest absolute bond-length deviation from reference values was recorded per molecule. Bonds for which no reference value was available ($0.75\%$) were omitted. The same analysis was carried out for molecules from the PubChemQC dataset containing the same atom types as QMugs, in order to obtain a comparable set with respect to the present atom types. The PubChemQC set ($3,834,382$ conformations with reference bond lengths) showed a deviation of $0.06$~$\pm$~\SI{0.04}{\angstrom} (median $\pm$ $1$ standard deviation), whereas the QMugs dataset ($2,004,003$ conformations with reference bond lengths) showed a deviation of $0.07$~$\pm$~\SI{0.03}{\angstrom}. Based on the observed distribution of bond-length deviations from experimental reference values (Figure~S1 in ESI) and manual investigation of example structures, \SI{0.2}{\angstrom} was determined to be a suitable threshold for a conformation to be removed from the dataset, which included $6,131$ ($0.31$\%) examples.

\subsubsection*{Molecular graph isomorphism}
It was investigated whether atom connectivity could be reconstructed after removing bond information from the generated SDFs. To this end, molecular graphs constructed exclusively from atom positions and types were compared to those obtained using the original atom connectivity (see ESI for details). $1,568$ ($0.08$\%) conformations for which the resulting molecular graphs were non-isomorphic failed this test.

\subsubsection*{Deviation of triple bonds from linear geometry}
The deviation of triple bonds from their ideal linear geometry was examined. In this investigation, ring triple bonds were not considered owing to routinely-occurring deviations from linear geometry in systems with high ring strain~\cite{bach2009ring}. The largest deviation from a \SI{180}{\degree} (linear) bond angle was recorded for each molecule containing at least one non-ring triple bond. The same analysis was performed on the PubChemQC dataset~\cite{nakata2017pubchemqc}. Triple-bond-containing molecules from PubchemQC and QMugs ($273,320$ and $165,101$ samples, respectively) show deviations of $1.38$~$\pm$~\SI{1.46}{\degree} (median $\pm$ $1$ standard deviation), and $1.46$~$\pm$~\SI{2.13}{\degree}, respectively. Based on the observed distribution of triple bond angles from a linear geometry (Figure~S2 in ESI) and manual inspection of structures, a \SI{10}{\degree} deviation was identified as a suitable threshold. $1,147$ ($0.06$\%) conformations failed this test.

\subsubsection*{Deviation of aromatic rings from planar geometry}
The planarity of carbon-containing aromatic rings was also investigated. For each molecule containing aromatic carbon atoms, the largest dihedral angle between the two planes spanned by each aromatic carbon atom and its three neighbors was recorded (see ESI for details). The same analysis was performed on the PubChemQC dataset~\cite{nakata2017pubchemqc}. Molecules from PubchemQC and QMugs ($2,391,589$ and $1,950,929$ conformations with aromatic carbons, respectively) showed median dihedral angles (~$\pm$~$1$ standard deviation) of $1.70$~$\pm$~\SI{1.85}{\degree} and $2.99$~$\pm$~\SI{2.20}{\degree}, respectively. Based on the observed distribution of dihedral angles from planar geometries (Figure~S3 in ESI) and manual inspection of structures, $2,769$ ($0.14$\%) conformations with aromatic carbon dihedral angles above \SI{15}{\degree} were discarded.

\subsection*{Further geometrical assessment}

\begin{figure}[ht]
\centering
\includegraphics[width=\linewidth]{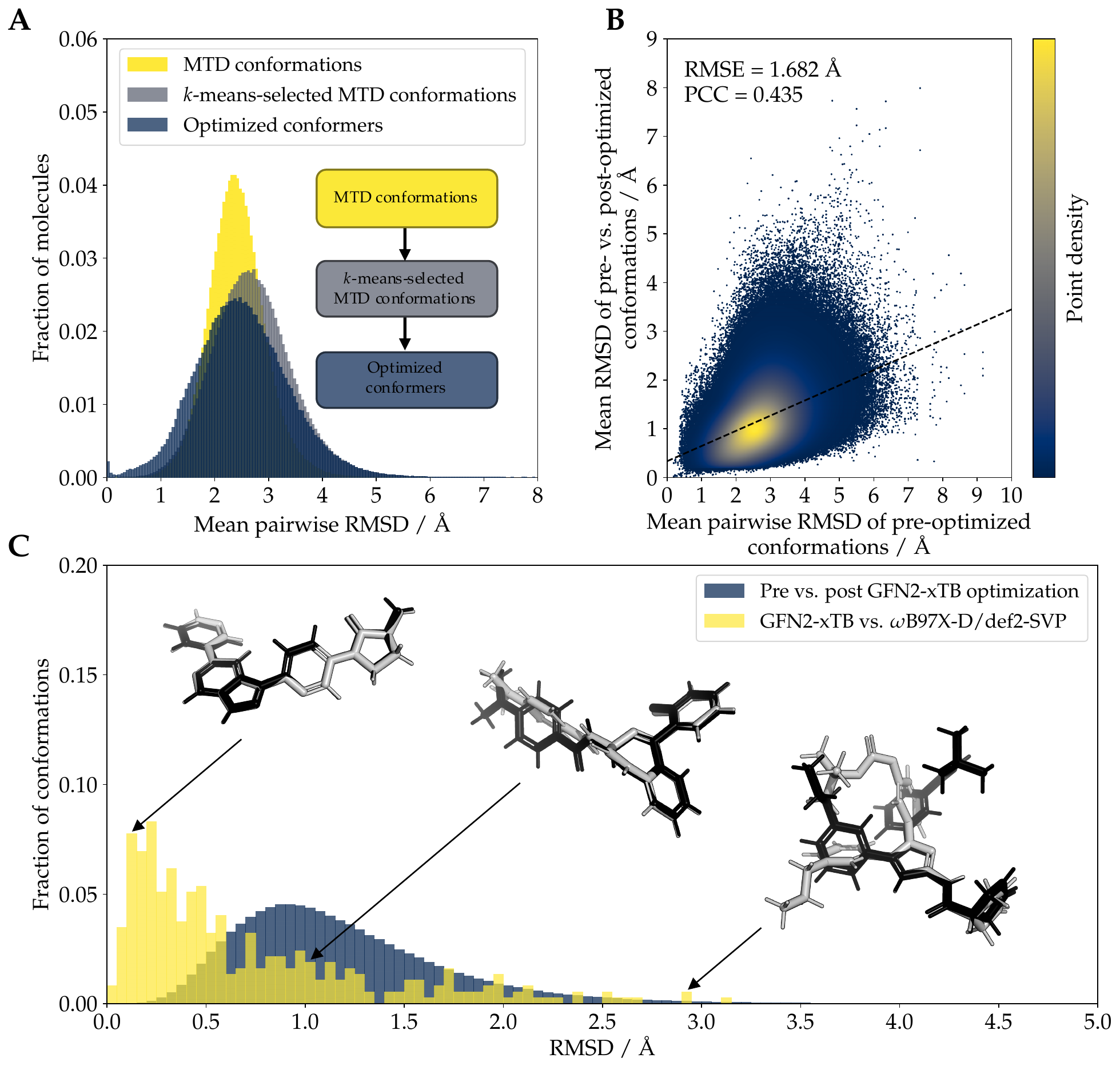}
\caption{
    (\textbf{A})~Distributions of mean pairwise RMSD of atom positions between conformations of each molecule in the QMugs dataset at different stages along the pipeline. While the $k$-means sampling process selects conformations that are, on average, more geometrically diverse than the average pair of structures generated by MTD simulations, geometry optimization reduces the geometrical diversity between the optimized conformers.
    (\textbf{B})~Change in atom positions during geometry optimization vs. mean pairwise RMSD of conformations before optimization. Molecules with initially more diverse conformations displayed a greater change in atom positions than those with initially less diverse conformations. 
    (\textbf{C})~Distribution of RMSD of structures prior to and after optimization with the semi-empirical GFN2-xTB method, and of structures optimized with the same approach vs. with $\omega$B97X-D/def2-SVP. The structures of three molecules with varying differences between the two methods are shown as illustrative examples (black and gray correspond to GFN2-xTB and $\omega$B97X-D/def2-SVP-optimized structures, respectively). For illustrative purposes, the example molecules are aligned on their substructures.
}
\label{fig:geometry_validation}
\end{figure}

The changes in the molecular geometries along the applied pipeline were examined in order to evaluate the effects of the applied steps. Figure~\ref{fig:geometry_validation}A shows the mean pairwise RMSD of atom positions between the conformations of each molecule at different steps along the pipeline. Conformations sampled during MTD simulations show a mean pairwise RMSD of $2.40$ $\pm$ $0.52$ \si{\angstrom} (median $\pm$ $1$ standard deviation). The $k$-means clustering procedure accomplishes the envisaged task of sampling conformations with higher geometric diversity ($2.67$ $\pm$ $0.74$ \si{\angstrom}). During the geometry optimization process, conformational diversity decreases ($2.48$ $\pm$ $0.86$ \si{\angstrom}). Unsurprisingly, for some molecules featuring rigid structures, conformations tend to converge toward the same energy minimum ($0.09$ \% of molecules show a mean pairwise RMSD $<$~\SI{0.01}{\angstrom} between their optimized conformers).

The degree to which the molecular geometries changed during the final optimization step was further analyzed. Molecules with initially more diverse conformations (higher mean pairwise RMSD of pre-optimized conformations) were shown to undergo a greater chance in atom positions (mean RMSD of pre- vs. post-optimized conformations) during optimization with the GFN2-xTB method (Figure~\ref{fig:geometry_validation}B). The observed heteroscedastic behavior of these two properties indicates that while the mean RMSD of pre- vs. post-optimized conformations tends to increase with higher mean pairwise RMSD of pre-optimized conformations, its variance also increases.

Finally, the suitability of GFN2-xTB as a lower-cost surrogate for DFT-level geometry optimization (Figure~\ref{fig:geometry_validation}C) was confirmed. $500$ randomly-chosen structures prior to semi-empirical geometry optimization from the QMugs dataset were further subjected to DFT-level geometry optimization ($\omega$B97X-D/def2-SVP), discarding structures that could not be converged in 100 iterations or with the computational resources described in the ESI. The RMSDs between the structures independently optimized at both levels of theory were then measured. The pairs of structures show RMSDs of $0.47$ $\pm$ $0.63$ \si{\angstrom} (median $\pm$ $1$ standard deviation), indicating that the chosen semi-empirical method obtains similar geometries to those obtained with more expensive first-principle calculations. Large RMSDs in some example pairs (Figure~\ref{fig:geometry_validation}C) could be interpreted as indicative of convergence to distinct local minima. 

For $2,067$ molecules, their individual conformations have different SMILES describing two different \textit{(E)/(Z)} isomers. Those structures are either $\alpha$-$\beta$-unsaturated ketones, $\alpha$-$\beta$-unsaturated nitriles, imine, or azo compounds, for which isomerization might be plausible~\cite{goulet2017electrocatalytic, roca2014dft}. In part due to the applied washing procedure, $17,176$ molecules can be represented with a SMILES string that is shared with at least one other ChEMBL-ID.

\subsection*{Validation of single-point properties}

\begin{figure}[ht]
\centering
\includegraphics[width=\linewidth]{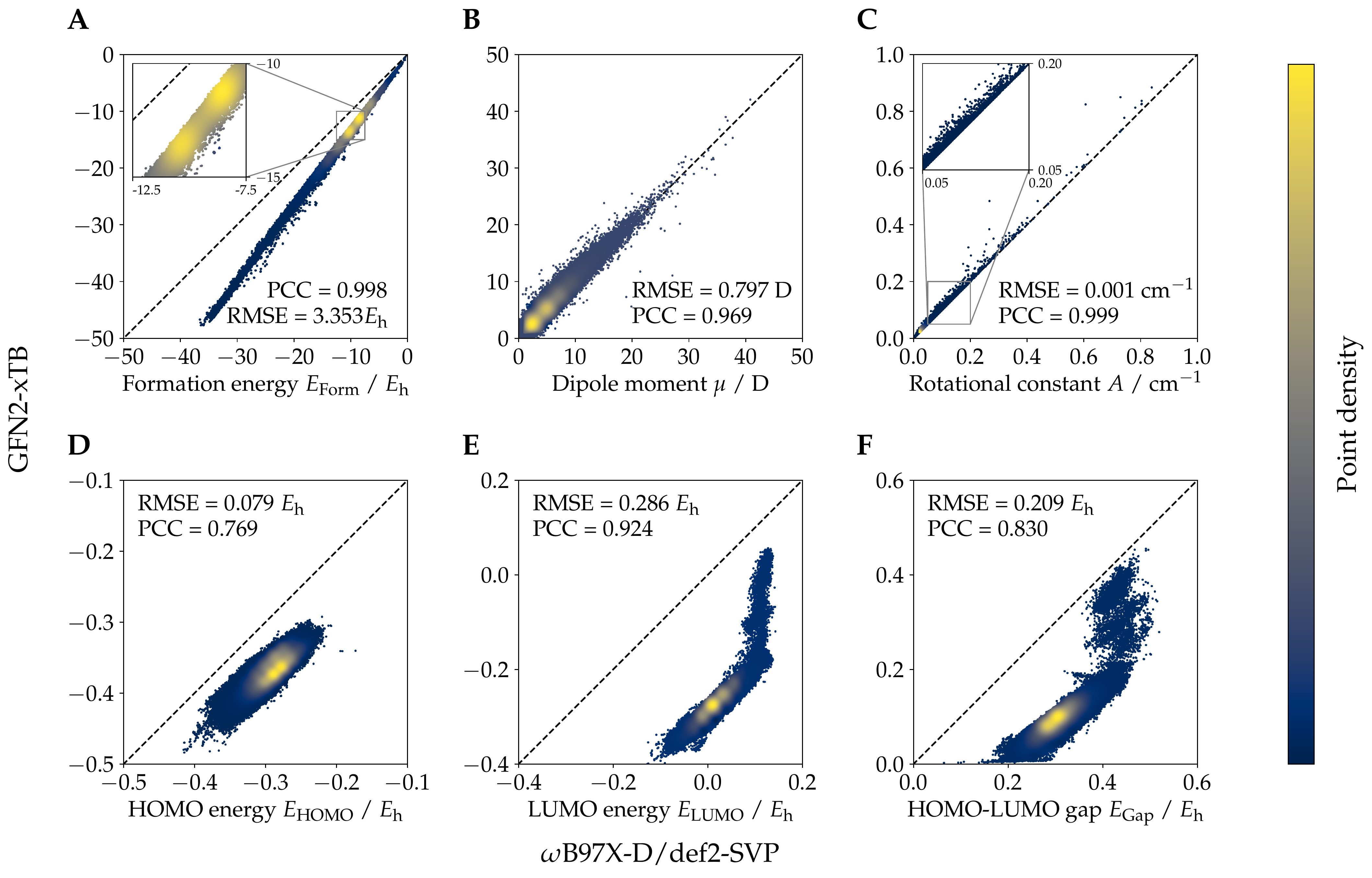}
\caption{Comparison of molecular properties computed at the two levels of theory considered herein (GFN2-xTB, $\omega$B97X-D/def2-SVP) for the molecules contained in QMugs. The molecular formation energy $E_{\mathrm{form}}$ in (\textbf{A}) was calculated by subtracting the atomic $U_{\mathrm{Atom}}$ contributions from the  total molecular energies $U_{RT}$. Only the rotational constants $A$ are shown in (\textbf{C}) as their $B$ and $C$ counterparts showed highly similar values. $22$ conformations of small molecules show very large rotational constants and are not shown. RMSE and PCC for rotational constant $A$ are $845.834$ cm$^{-1}$ and $0.091$ respectively, if those structures are included. Abbreviations: RMSE, root mean squared error; PCC, Pearson's correlation coefficient.}
\label{fig:delta_molecular_props}
\end{figure}

\begin{figure}[ht]
\centering
\includegraphics[width=\linewidth]{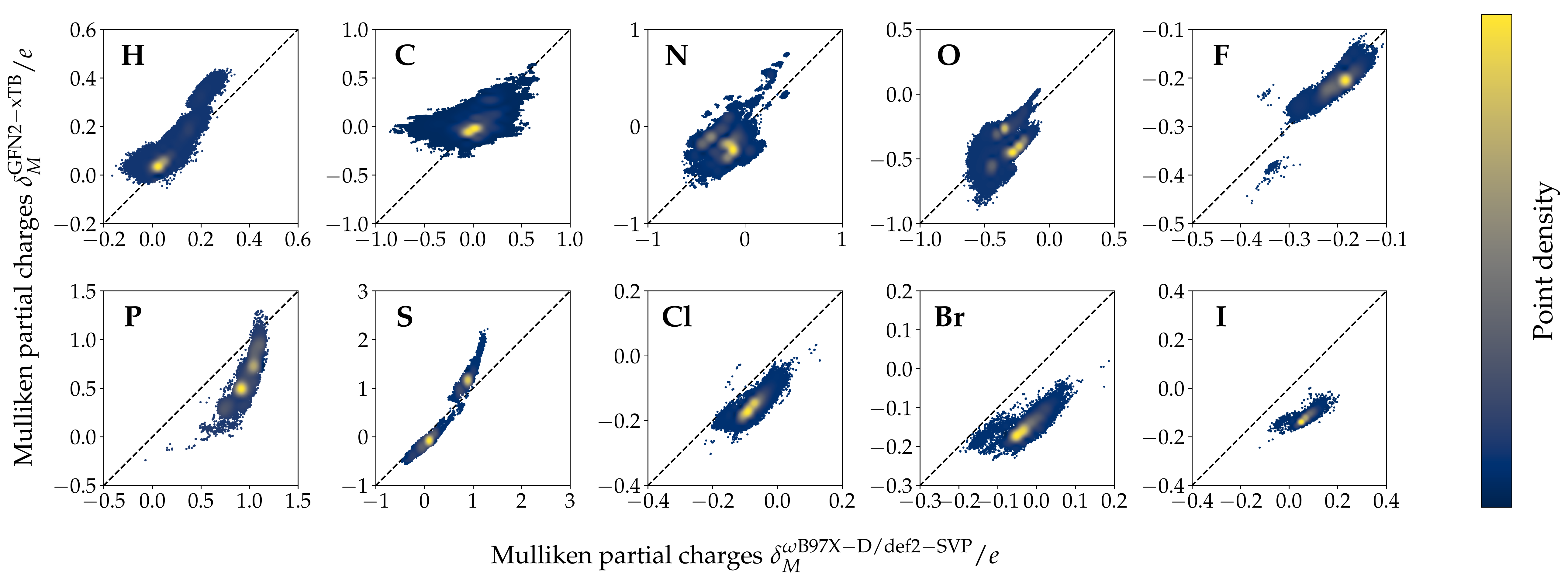}
\caption{Atom-type-specific partial charge correlations (GFN2-xTB, $\omega$B97X-D/def2-SVP) for the QMugs dataset (see ESI Table~1 for additional metrics)}
\label{fig:partial}
\end{figure}

\begin{figure}[ht]
\centering
\includegraphics[width=\linewidth]{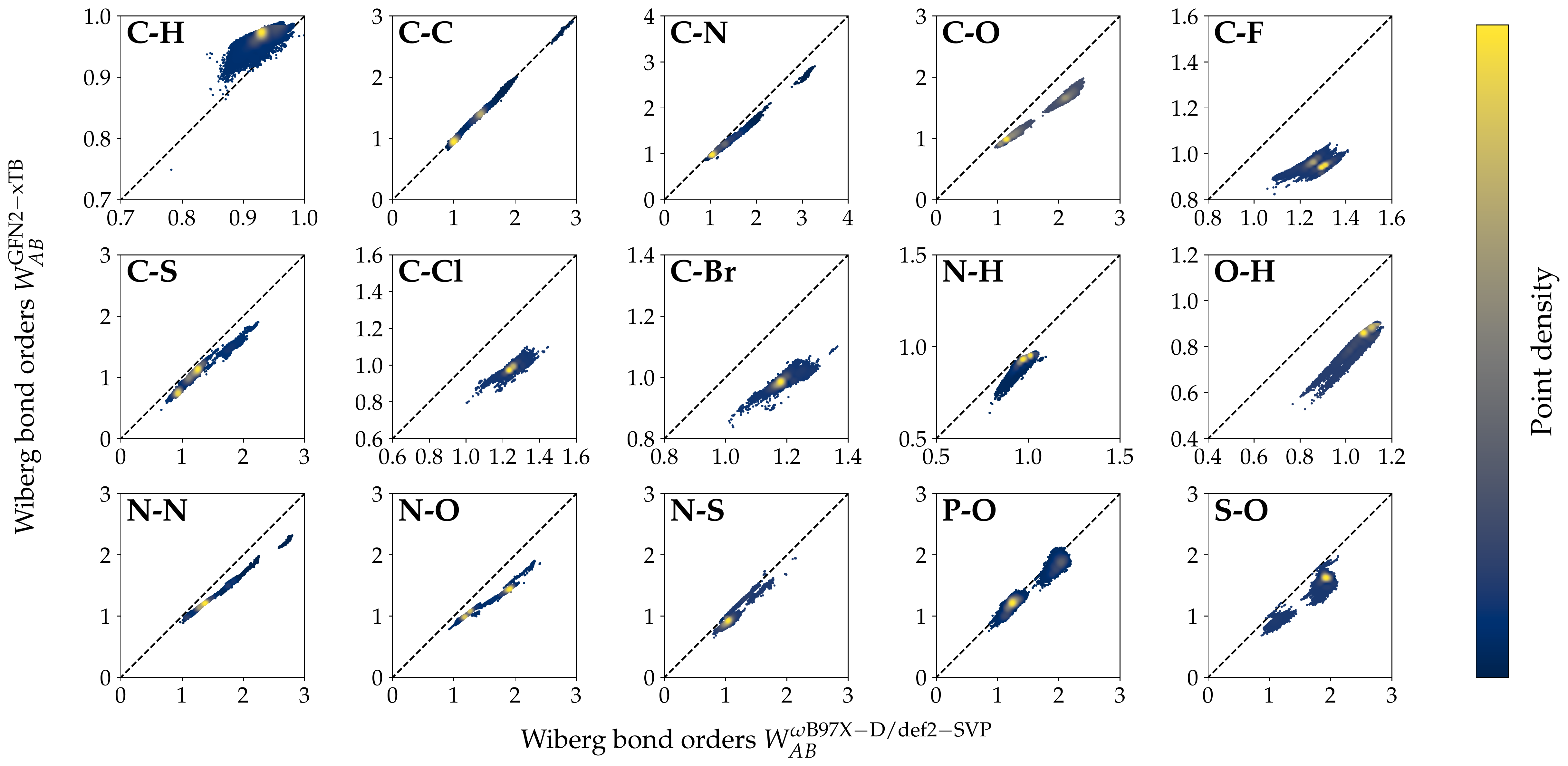}
\caption{Comparison of Wiberg bond orders between GFN2-xTB and $\omega$B97X-D/def2-SVP for the 15 most frequently occurring bond types in the QMugs dataset. The latter level of theory uses L\"owdin-orthogonalization. See ESI Table~2 for additional metrics. For bond types which occurred $>1$M times in the dataset, a randomly chosen sample of $1$M bonds is plotted.}
\label{fig:bond}
\end{figure}

To validate the general agreement between the two methods employed in this work, the correlation between a series of single-point properties computed on both levels of theory was analyzed. Both global, molecular (Figure~\ref{fig:delta_molecular_props}) and local, atomic/bond properties (Figures~\ref{fig:partial}~\&~\ref{fig:bond}) were considered. All single-point molecular properties showed a high degree of correlation. Formation energies $E_\mathrm{Form}$ (Figure~\ref{fig:delta_molecular_props}A), which were obtained by subtracting atomic energies $E_{\mathrm{Atom}}$ from total internal energies $U_{RT}$, show a Pearson correlation coefficient (PCC) of $0.998$. Dipole moments $\mu$ and rotational constants $A$ (excl. $22$ small structures with very high rotational constants; Figure~\ref{fig:delta_molecular_props}B,C) display PCCs of $0.969$ and $0.999$, respectively. Orbital energies, namely the energies for highest occupied (HOMO)  $E_{\mathrm{HOMO}}$ and lowest unoccupied molecular orbitals (LUMO) $E_{\mathrm{LUMO}}$ and HOMO-LUMO gaps $E_{\mathrm{Gap}}$ show PCCs of $0.769$, $0.924$ and $0.830$, respectively (Figures~\ref{fig:delta_molecular_props}~D,~E~\&~F). The observed PCCs for all six single-point molecular properties indicate good agreement between the two methods. Atom-type-specific partial charges for the 10 atom-types in QMugs (Figure~\ref{fig:partial}, ESI Table~1) as well as the 15 most abundant covalent bonds orders (Figure~\ref{fig:bond}, ESI Table~2) also show high correlations between the two methods used herein. Regarding partial charges, $7$ out of the $10$ atom types considered in QMugs were observed to have PCCs $>0.8$, with the remaining carbon, nitrogen, and oxygen atom-types resulting in lower PCCs of $0.574$, $0.124$, and $0.274$, respectively. Regarding bond orders, $10$ out of the $15$ show PCCs $>0.9$ and $14$ out of $15$ display PCCs $>0.75$ (see ESI Table~2 for additional metrics). Notably only carbon-fluorine bonds display a larger discrepancy between both levels of theory, with an observed PCC of $0.153$. The observed correlations in both molecular and atomic single-point properties between GFN2-xTB and $\omega$B97X-D/def2-SVP confirm the suitability of the former method as a computationally affordable starting point for $\Delta$-learning of DFT-level properties.

\section*{Usage Notes}

All data files can be accessed via any modern web browser, and can be programmatically downloaded using the provided instructions in the archive's readme. The provided SDFs can be processed using standard cheminformatics software (\textit{e.g.}, RDKit~\cite{rdkit}, KNIME~\cite{knime}), and wave function files using the Psi4~\cite{smith2020psi4} software package or directly using Numpy~\cite{harris2020array}. 

\section*{Code availability}

All analyses were supported by the Python programming language (version 3.7.7) and its scientific software stack~\cite{harris2020array}. Molecular conformations were generated using RDKit~\cite{rdkit} (version 2020.03.3) and GFN2-xTB~\cite{grimme2017robust, bannwarth2019gfn2, pracht2019robust, grimme2019exploration, bannwarth2020extended} (version 6.3.1). All quantum mechanical calculations were carried out with Psi4~\cite{smith2020psi4} (version 1.3.2). Molecular structure visualizations were created using PyMol~\cite{PyMOL} (version 2.3.5) and ChemDraw (version 19.1.1.32). The rclone (\url{https://rclone.org}) WebDAV client was used for all data uploading purposes.

\section*{Acknowledgements} 
We thank Dr. Jan A. Hiss and Dr. Petra Schneider for their valuable insights and discussions. Special thanks to Andreas la Roi from the ETH Research Collection team and to the ETH cluster team for their technical support throughout this project. This work was financially supported by the ETH RETHINK initiative, the Swiss National Science Foundation (grant no. 205321\_182176), and Boehringer Ingelheim Pharma GmbH \& Co. KG. C.I. acknowledges  support from the Scholarship Fund of the Swiss Chemical Industry.\\

\section*{Author contributions statement}
\textbf{Clemens Isert}: Methodology, Formal Analyses, Writing. \textbf{Kenneth Atz}: Methodology, Formal Analyses, Writing. \textbf{José Jiménez-Luna}: Conceptualization, Methodology, Writing. \textbf{Gisbert Schneider}: Supervision, Writing.


\section*{Competing interests}
G.S. is a cofounder of inSili.com LLC, Zurich, and a consultant to the pharmaceutical industry.

\bibliography{references}

\section*{Figures \& Tables}

\begin{table}[ht]
\caption{Descriptive statistics of the dataset reported herein in the context of other  DFT-level molecular datasets and the information provided by each. The number of molecules for PubChemQC corresponds to that available on the website of the project.~\cite{pubchemqc_website} Heavy atom averages are weighted by the number of conformations.}
\label{tbl:desc}
\centering
\resizebox{\textwidth}{!}{%
\begin{tabular}{@{}l>{\raggedleft\arraybackslash}p{2cm}>{\raggedleft\arraybackslash}p{2cm}>{\raggedleft\arraybackslash}p{2cm}p{5cm}P{2cm}P{2cm}p{3cm}@{}}
\toprule
\textbf{Dataset} &
  \textbf{Unique compounds} &
  \textbf{Total conformations} &
  \textbf{Heavy atoms max (mean)} &
  \textbf{Method} &
  \textbf{$\Delta$-learning possible} &
  \textbf{Wave functions} \\ \midrule
QM9       & $133,885$  & $133,885$   & $9$~~~($8.8$) & B3LYP/6-31G(2df,p)               & \xmark &  \xmark \\
ANI-1     & $57,462$   & $22,057,374$ & $8$~~~($7.1$) & $\omega$B97X/6-31G(d)              & \xmark  & \xmark \\
PubChemQC & $3,982,436$ & $3,982,436$  & $51$ ($14.1$) & B3LYP/6-31G(d)              & \xmark  & \xmark \\
QMugs     & $665,911$  & $1,992,984$  & $100$ ($30.6$) & GFN2-xTB + $\omega$B97X-D/def2-SVP & \cmark  & \cmark \\ \bottomrule
\end{tabular}
}
\end{table}

\begin{table}[ht]
\footnotesize
\caption{Calculated properties as stored in the SDFs of the QMugs data collection. Abbreviations: a.u., atomic units; vib., vibrational; rot., rotational; transl., translational. Properties that enable $\Delta$ machine learning are labelled with $\blacklozenge$.}
\label{tbl:properties}
\centering
\begin{tabular}{llllll}
\toprule
\textbf{Property}                                        & \textbf{Symbol}                               & \textbf{Unit}           & \textbf{Key}  & $\Delta$-ML        \\  \midrule
ChEMBL identifier                                               &         -                      &           -              & \texttt{CHEMBL\_ID} &                \\
Conformer identifier                                                 &     -                                          &                    -     & \texttt{CONF\_ID}      &              \\
Total energy                                             & $U_{RT}$            & \si{\hartree}                      & \texttt{GFN2:TOTAL\_ENERGY} & $\blacklozenge$                  \\
Internal atomic energy                             & $E_\mathrm{Atom}$                             & \si{\hartree}                      & \texttt{GFN2:ATOMIC\_ENERGY} &       \\
Formation energy                             & $E_\mathrm{Form}$           & \si{\hartree}                      & \texttt{GFN2:FORMATION\_ENERGY}   & $\blacklozenge$          \\
Total enthalpy                                           & $H_{RT}$                                      & \si{\hartree}                      & \texttt{GFN2:TOTAL\_ENTHALPY}    &        \\
Total free energy                                        & $G_{RT}$                                      & \si{\hartree}                      & \texttt{GFN2:TOTAL\_FREE\_ENERGY} &        \\
Dipole ($x$, $y$, $z$, total)                  & $\mu$                                         & D                       & \texttt{GFN2:DIPOLE}      & $\blacklozenge$                    \\
Quadrupole ($xx$, $xy$, $yy$, $xz$, $yz$, $zz$) & $Q_{ij}$                                      & D \si{\angstrom}                 & \texttt{GFN2:QUADRUPOLE}      &           \\
Rotational constants ($A$, $B$, $C$)           & $A$, $B$, $C$                              & \si{\centi\meter}$^{-1}$                     & \texttt{GFN2:ROT\_RONSTANTS}    & $\blacklozenge$              \\
Enthalpy (vib., rot., transl., total)          & $\Delta H$                                    & cal mol$^{-1}$          & \texttt{GFN2:ENTHALPY}     &               \\
Heat capacity (vib., rot., transl., total)    & $C_{V}$                                       & cal K$^{-1}$ mol$^{-1}$ & \texttt{GFN2:HEAT\_CAPACITY}  &             \\
Entropy (vib., rot., transl., and total)       & $\Delta S$                                    & cal K$^{-1}$ mol$^{-1}$ & \texttt{GFN2:ENTROPY}     &                \\
HOMO energy                                              & $E_\mathrm{HOMO}$                             & \si{\hartree}                      & \texttt{GFN2:HOMO\_ENERGY}       & $\blacklozenge$             \\
LUMO energy                                              & $E_\mathrm{LUMO}$                             & \si{\hartree}                      & \texttt{GFN2:LUMO\_ENERGY}     & $\blacklozenge$           \\
HOMO-LUMO gap                                            & $E_\mathrm{Gap}$                              & \si{\hartree}                      & \texttt{GFN2:HOMO\_LUMO\_GAP}  & $\blacklozenge$               \\
Fermi level                                              & $E_{\mathrm{Fermi}}$                                   & \si{\hartree}                      & \texttt{GFN2:FERMI\_LEVEL}   &             \\
Mulliken partial charges                                 & $\delta_{M}$                                  & \si{\elementarycharge}                       & \texttt{GFN2:MULLIKEN\_CHARGES} & $\blacklozenge$              \\
Covalent coordination number                           & $N_{\textrm{coord}}$                  & -                   &\texttt{GFN2:COVALENT\_COORDINATION\_NUMBER}  & \\
Molecular dispersion coefficient                           & $C_6$                                                & a.u.                            &\texttt{GFN2:DISPERSION\_COEFFICIENT\_MOLECULAR} & \\
Atomic dispersion coefficients                         & $C_6$                                                & a.u.                                &\texttt{GFN2:DISPERSION\_COEFFICIENT\_ATOMIC} & \\
Molecular polarizability                           & $\alpha(0)$                                                 & a.u.                                        &\texttt{GFN2:POLARIZABILITY\_MOLECULAR} &  \\
Atomic polarizabilities                            & $\alpha(0)$                                                 & a.u.                                        &\texttt{GFN2:POLARIZABILITY\_ATOMIC} &  \\
Wiberg bond orders                                    & $M_{AB}$                                 &             -            &\texttt{GFN2:WIBERG\_BOND\_ORDER}     & $\blacklozenge$            \\
Total Wiberg bond orders                               & $\sum_{A (A \neq B)} M_{AB}$                      &       -                  &\texttt{GFN2:TOTAL\_WIBERG\_BOND\_ORDER}  & $\blacklozenge$              \\
Total energy                                             & $U_{RT}$                                      & \si{\hartree}                      & \texttt{DFT:TOTAL\_ENERGY}     & $\blacklozenge$             \\
Total internal atomic energy                             & $E_\mathrm{Atom}$                             & \si{\hartree}                      & \texttt{DFT:ATOMIC\_ENERGY}    &         \\
Formation energy                             & $E_\mathrm{Form}$                             & \si{\hartree}                      & \texttt{DFT:FORMATION\_ENERGY}  & $\blacklozenge$           \\
Electrostatic potential                                  & $V_{ESP}$                                     & \si{\volt}                       & \texttt{DFT:ESP\_AT\_NUCLEI}   &         \\
L\"owdin partial charges                                 & $\delta_{L}$                                  & \si{\elementarycharge}                       & \texttt{DFT:LOWDIN\_CHARGES}     &       \\
Mulliken partial charges                                 & $\delta_{M}$                                  & \si{\elementarycharge}                       & \texttt{DFT:MULLIKEN\_CHARGES}   & $\blacklozenge$           \\
Rotational constants  ($A$, $B$, $C$)          & $A$, $B$, $C$                              & \si{\centi\meter}$^{-1}$                     & \texttt{DFT:ROT\_CONSTANTS}   & $\blacklozenge$              \\
Dipole ($x$, $y$, $z$, total)                  & $\mu$                                         & D                       & \texttt{DFT:DIPOLE}                     \\
Exchange correlation energy                              & $\hat{V}_{eN}$                                & \si{\hartree}                      & \texttt{DFT:XC\_ENERGY}   &              \\
Nuclear repulsion energy                                 & $\hat{V}_{eN}$                                & \si{\hartree}                      & \texttt{DFT:NUCLEAR\_REPULSION\_ENERGY} & \\
One-electron energy                               & $\hat{T}_{e}$                                 & \si{\hartree}                      & \texttt{DFT:ONE\_ELECTRON\_ENERGY}   &    \\
Two-electron energy                                      & $\hat{V}_{ee}$                                & \si{\hartree}                      & \texttt{DFT:TWO\_ELECTRON\_ENERGY} &     \\
HOMO energy                                              & $E_\mathrm{HOMO}$                             & \si{\hartree}                      & \texttt{DFT:HOMO\_ENERGY}    & $\blacklozenge$               \\
LUMO energy                                              & $E_\mathrm{LUMO}$                             & \si{\hartree}                      & \texttt{DFT:LUMO\_ENERGY}    & $\blacklozenge$               \\
HOMO-LUMO gap                                            & $E_\mathrm{Gap}$                              & \si{\hartree}                      & \texttt{DFT:HOMO\_LUMO\_GAP}   & $\blacklozenge$             \\
Mayer bond orders                                        & $M_{AB}$                                 &                -         & \texttt{DFT:MAYER\_BOND\_ORDER}   &           \\
Wiberg-L\"owdin bond orders                              & $W_{AB}$                                 &            -             & \texttt{DFT:WIBERG\_LOWDIN\_BOND\_ORDER}  & $\blacklozenge$      \\
Total Mayer bond orders                              & $\sum_{A (A \neq B)} M_{AB}$                      &          -               & \texttt{DFT:TOTAL\_MAYER\_BOND\_ORDER}      &      \\
Total Wiberg-L\"owdin bond orders                          & $\sum_{A (A \neq B)} W_{AB}$        &      -                   & \texttt{DFT:TOTAL\_WIBERG\_LOWDIN\_BOND\_ORDER}  & $\blacklozenge$     \\ \bottomrule
\end{tabular}
\end{table}

\begin{table}[ht]
\caption{Calculated molecular properties stored in the wave function files provided in the QMugs data collection. Mayer and Wiberg-L\"owdin bond orders included here represent a superset of the bond orders in the SDFs which additionally comprise bond orders for non-covalent bonds.}
\label{tbl:wfn}
\centering
\begin{tabular}{llll}
\toprule
\textbf{Property}                                        & \textbf{Symbol}                               & \textbf{Key}                            \\ \midrule
Alpha density matrix                                     & $\mathrm{D}_{\alpha}$                         & \texttt{matrix, Ca}                              \\
Beta density matrix                                      & $\mathrm{D}_{\beta}$                          & \texttt{matrix, Cb}                              \\
Alpha orbitals                                           & $\mathrm{C}_{\alpha}$                         & \texttt{matrix, Da}                              \\
Beta orbitals                                            & $\mathrm{C}_{\beta}$                          & \texttt{matrix, Db}                              \\
Atomic-orbital-to-symmetry-orbital transformer           & $\mathrm{C}_{\mathrm{AOTOSO}}$                & \texttt{matrix, aotoso}                          \\
Mayer bond orders                                        & $M_{AB}$                                      & \texttt{MAYER\_INDICES}                          \\
Wiberg-L\"owdin bond orders                              & $W_{AB}$                                      & \texttt{WIBERG\_LOWDIN\_INDICES}                 \\ \bottomrule
\end{tabular}
\end{table}

\end{document}


\renewcommand{\thefigure}{S\arabic{figure}}

\maketitle


\section{Overlap with other datasets}
The overlap between the compounds included in the QMugs dataset with those included in other datasets featuring DFT-properties, namely QM9~\cite{ramakrishnan2014quantum}, PubchemQC~\cite{nakata2017pubchemqc} and ANI-1~\cite{smith2017anid}, was investigated. Overlap was computed based on the respective compounds' InChI~\cite{heller2015inchi} strings. For each dataset, the molecules were converted from their original formats (QMugs: \texttt{.sdf}, QM9: \texttt{.xyz}, PubChemQC: \texttt{.mol}, ANI-1: SMILES~\cite{weininger1988smiles}, as obtained with the \texttt{pyanitools} module included in the ANI-1 repository) to InChIs using Openbabel~\cite{o2011open, openbabel} (version 3.1.1). Molecules that could not be successfully converted were skipped. The Venn diagram (Fig. 4, main article) was constructed using the \texttt{pyvenn}~\cite{pyvenn} package. Note that due to different washing procedures, the same compound identifier may refer to different InChI strings in different datasets (\textit{e.g.}, protonated and unprotonated).

\section{ChEMBL SQL query}
A locally-downloaded MySQL instance of the ChEMBL27 database~\cite{mendez2019chembl} (version 8.0.19) was queried~\cite{mysql}. Our data extraction procedure is performed in two steps: First, a list of biological targets was extracted. Second, compounds for which a specific activity towards any of these targets was annotated were extracted. Both steps are described in the following. 

\subsection{Extraction of targets}
"Single-protein" targets were selected, for which \texttt{activity.standard\_types} were reported either as $\mathrm{IC}_{50}$, $\mathrm{EC}_{50}$, $K_i$, $K$, $K_b$, $K_a$, $K_d$, $K_e$, or $K_m$ and in units of nM, or as a set of other \texttt{activity.standard\_types} (\textit{e.g.,} $-\log K$) and without units. The following types of annotations were excluded: Annotations with an \texttt{assay.relationship\_type} other than "homologous" or "direct", annotations with an \texttt{assay.} \texttt{confidence\_score} below $7$, annotations with assay descriptions for mutant species, annotations for potential activity duplicates, annotations with assay data validity comments other than "Manually validated" or "null", and  annotations with \texttt{activity.activity\_comments} indicating missing data such as "Not Determined" or "Not tested". Only targets which had annotations for $10$ or more unique compounds (as identified by their \texttt{activity.molregno} identifier) after this data extraction process were kept. 

\subsection{Extraction of compounds}
For each of the targets extracted in the previous steps, we queried the SQL database for activity annotations. Annotations for which the \texttt{activity.standard\_value} or the \texttt{activity.standard\_unit} was missing were discarded. All \texttt{activity.standard\_values} for annotations were converted to negative decadic logarithm scales (p$X$) and values below -3 on this scale were corrected to their absolute value. The  \texttt{activity.standard\_values} for annotations with activity comments denoting inactivity were set to $3$ on this log-scale. Annotations where the converted \texttt{activity.standard\_value} lay outside the range of $3$-$12$ were excluded. For each of the compounds that remained, we extracted the ChEMBL-ID and the canonical SMILES.

\section{SMILES filtering}
The molecules extracted from the ChEMBL database were filtered to exclude the following SMARTS patterns using RDKit~\cite{rdkit} (version 2020.03.3.0): \\

\texttt{B}, \texttt{[*]$\sim$P($\sim$[*])($\sim$[*])($\sim$[*])$\sim$[*]}, \texttt{[*]$\sim$S($\sim$[*])($\sim$[*])($\sim$[*])$\sim$[*]}, \texttt{[S+$^\wedge$3]}, \texttt{[P+$^\wedge$3]}, \newline 
\texttt{[*]$\sim$[F,Cl,Br,I]$\sim$[*]}, \texttt{[Be]}, \texttt{[Na]}, \texttt{[Al]}, \texttt{[Si]}, \texttt{[Sc]}, \texttt{[Ti]}, \texttt{[V]}, \texttt{[Cr]}, \texttt{[Mn]}, \texttt{[Fe]}, \texttt{[Co]}, \texttt{[Ni]}, \texttt{[Cu]}, \texttt{[Zn]}, \texttt{[Ga]}, \texttt{[Ge]}, \texttt{[As]}, \texttt{[Se]}, \texttt{[Rb]}, \texttt{[Sr]}, \texttt{[Y]}, \texttt{[Zr]}, \texttt{[Nb]}, \texttt{[Mo]}, \texttt{[Tc]}, \texttt{[Ru]}, \texttt{[Rh]}, \texttt{[Pd]}, \texttt{[Ag]}, \texttt{[Cd]}, \texttt{[In]}, \texttt{[Sn]}, \texttt{[Sb]}, \texttt{[Te]}, \texttt{[Cs]}, \texttt{[Ba]}, \texttt{[La]}, \texttt{[Ce]}, \texttt{[Pr]}, \texttt{[Nd]}, \texttt{[Pm]}, \texttt{[Sm]}, \texttt{[Eu]}, \texttt{[Gd]}, \texttt{[Tb]}, \texttt{[Dy]}, \texttt{[Ho]}, \texttt{[Er]}, \texttt{[Tm]}, \texttt{[Yb]}, \texttt{[Lu]}, \texttt{[Hf]}, \texttt{[Ta]}, \texttt{[W]}, \texttt{[Re]}, \texttt{[Os]}, \texttt{[Ir]}, \texttt{[Pt]}, \texttt{[Au]}, \texttt{[Hg]}, \texttt{[Pb]}, \texttt{[Bi]}, \texttt{[Fr]}
\\

Compounds containing these substructures were excluded as they were either incompatible with the used MMFF94s force field\cite{tosco2014bringing} (\textit{e.g.,} \texttt{B}), contained halogens bound to two neighbors (\textit{e.g.,} \texttt{[*]$\sim$[F,Cl,Br,I]$\sim$[*]}), or elements with reduced relevance for drug-like molecules.

\section{Computational resources}
For practical reasons, molecules whose DFT-calculations demanded computational resources that exceeded empirically determined limits, were discarded. These limits include a maximum processing power of $4$ CPU cores for $24$ h wall-time, up to $40$ GB of system memory, and $100$ GB of scratch space. For the single-point property calculations, this was the case for $9$ structures. Additionally, $19$ structures (GFN2-xTB) and $24$ structures (DFT) were discarded as the respective calculations could not be completed successfully.
\section{Optimized geometry sanity checks}
We performed four consecutive geometry checks (Sections~\ref{sec:bond_length_check} -- \ref{sec:planar_rings}) to filter out structures for which geometry optimization had converged to unrealistic conformations. Structures failing any of these tests were removed from the dataset. Note that stated numbers of investigated molecules for the QMugs dataset are larger than the size of the final dataset, as only molecules that passed all geometry checks were included in the final dataset. Note that the reported numbers of conformations failing the tests described in the following subsections do not add up to $10,986$ (the number of conformations removed from the dataset) since some conformations failed in multiple tests.

\subsection{Deviation of bond lengths from experimental reference values}
\label{sec:bond_length_check}

We searched for bond-length outliers in the optimized structures. As reference values, average bond-length values between atom types from ``Standard Reference Database, Computational Chemistry Comparison and Benchmark DataBase''~\cite{nist_database} were extracted. For each bond in the optimized structures, the absolute deviation from its respective reference value was calculated. For each conformation, the largest absolute deviation was recorded. If no reference value for a specific bond type between two atom types was found, the bond was not considered for further analysis. This was the case for $876,113$ ($0.75$\%) of all bonds in the investigated molecules from the QMugs dataset. 

\begin{figure}[H]
\centering
\includegraphics[width=\textwidth]{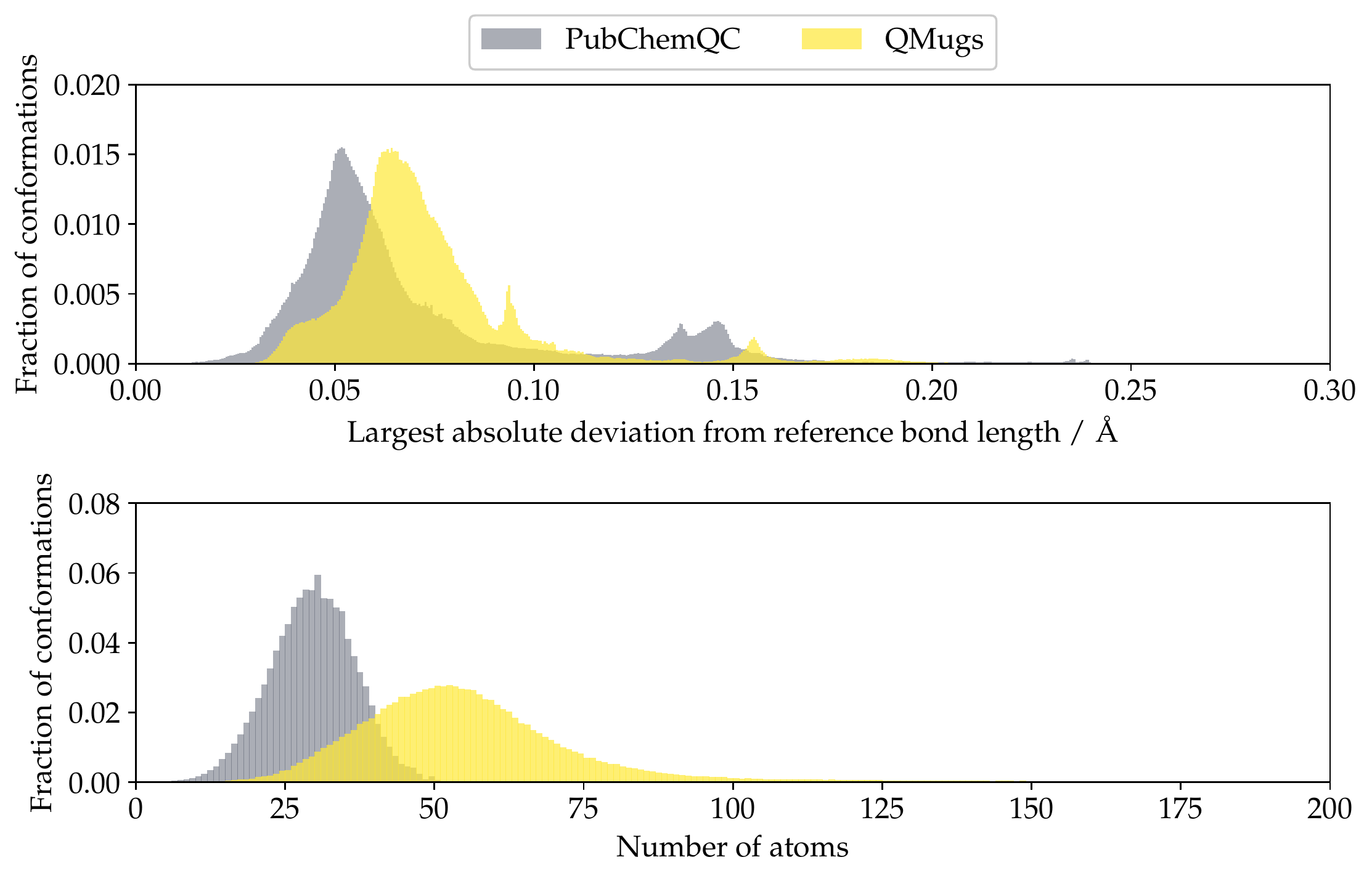}
\caption{
\textit{(Top)} Distribution of largest absolute bond-length deviation from experimental reference values per conformation (histogram bin size $5\times 10^{-4}$\si{\angstrom}). PubChemQC ($3,834,382$ conformations with reference bond lengths) shows a deviation of $0.0580$~$\pm$~\SI{0.0419}{\angstrom} (median $\pm$ $1$ standard deviation), whereas QMugs ($2,004,003$ conformations with reference bond lengths) exhibits a deviation of $0.0687$~$\pm$~\SI{0.0317}{\angstrom}. $8,492$ ($0.22$\%) and $926$ ($0.05$\%) conformations in the PubChemQC and QMugs sets, respectively, have higher deviations than \SI{0.30}{\angstrom} and are not shown. \textit{(Bottom)} Distribution of the total number of atoms per conformation in both datasets (histogram bin size $1$), showing that molecules in the QMugs sample are significantly larger on average. $765$ ($0.04$\%) conformations in the QMugs dataset have more than $200$ atoms and are not shown. Potentially-arising greater steric clashes in larger molecules may contribute to the slightly higher bond length deviations in the QMugs dataset, compared to the PubChemQC dataset.} 
\label{fig:bond_length_histograms}
\end{figure}

The same procedure was carried out on molecules from the PubChemQC dataset~\cite{nakata2017pubchemqc} containing the same atom types as the QMugs dataset (Figure~\ref{fig:bond_length_histograms}). The restriction of atom types was made in order to rule out effects of atom types on bond length deviations. This restriction removed $65,316$ ($1.64$\%) molecules from the assessment. Further $82,708$ ($2.08$\%) molecules which could not be successfully read with RDKit~\cite{rdkit} were ignored. Based on the distribution of bond-length deviations from experimental reference values, we discarded molecules above the \SI{0.2}{\angstrom} threshold. Based on manual inspection of structures at different bond length deviation values, this threshold appeared as a suitable, conservative threshold. $6,131$ ($0.31$\%) conformations did not pass this test.

\noindent

\subsection{Molecular graph isomorphism}
We investigated whether heavy-atom connectivity can be reconstructed when removing all bond information from the generated structure-data files (SDF) in the database. SDFs were converted to the .xyz file format (which does not contain bond information) using OpenBabel~\cite{o2011open, openbabel} (version 3.1.1). We then attempted to perform a conversion from .xyz to InChI~\cite{heller2015inchi}, or to SMILES~\cite{weininger1988smiles} upon failure of the former. If both were unsuccessful, we considered the graph isomorphism check as failed. If either succeeded, however, we compared the molecular graph of generated molecular strings (as read by RDKit) to the one originally obtained from the SDF (which includes bond information). The isomorphism of the molecular graphs was then checked using the NetworkX~\cite{SciPyProceedings_11} Python package (version 2.5), considering nodes (representing atoms) in each graph labelled with their respective atom types. We did not use bond types in the previous comparison due to observed high false negative rates related to mislabelled bonds in nitro and other functional groups with multiple resonance structures, among further reasons. $1,568$ ($0.08$\%) conformations failed this test. 

\subsection{Deviation of triple bonds from linear geometry}
\begin{figure}[H]
\centering
\includegraphics[width=\textwidth]{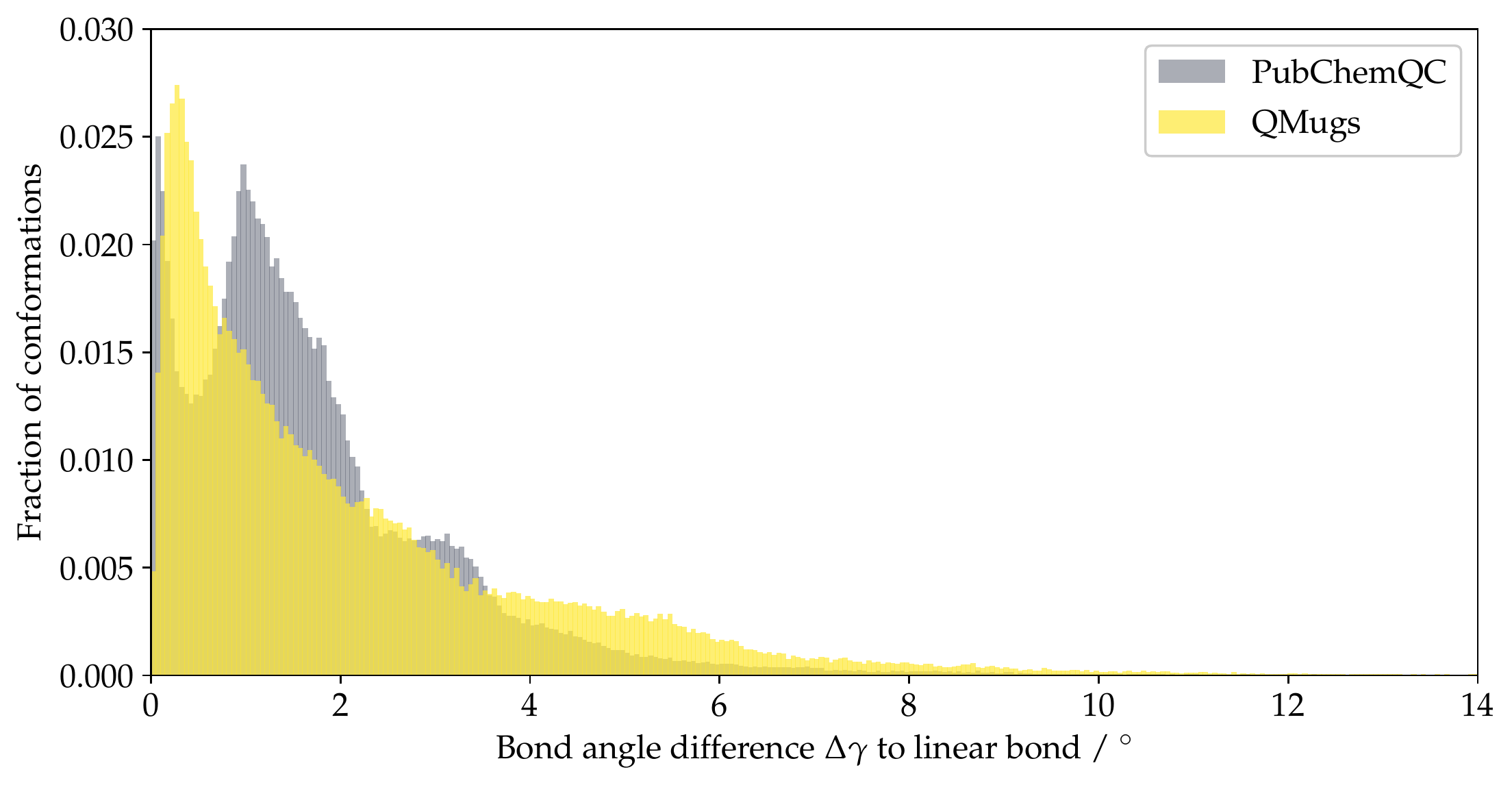}
\caption{Distribution of triple bond angle deviations from an ideal \SI{180}{\degree} angle (histogram bin size \SI{0.05}{\degree}). Triple bond-containing conformations from PubchemQC ($273,320$ conformations) and QMugs ($165,101$ conformations) show a deviation of $1.38$~$\pm$~\SI{1.46}{\degree} (median $\pm$ $1$ standard deviation), and $1.46$~$\pm$~\SI{2.13}{\degree}, respectively. $104$ conformations ($0.04$\%) in the PubChemQC and $179$ molecules ($0.11$\%) in the QMugs sample with higher deviations than \SI{14}{\degree} are not shown.} 
\label{fig:linear_triple_bonds}
\end{figure}

For each molecule containing a triple bond which is not part of a ring, the deviation of bond angles $\gamma$ from the ideal 180\textdegree ~(linear) geometry, denoted as $\Delta \gamma$, was assessed. For each non-terminal atom in a non-ring triple bond, the angle between the bonds to its two neighbors was computed. The largest deviation $\Delta \gamma$ from a perfectly linear triple bond was recorded per molecule. We limit this investigation to triple bonds outside a ring as triple bonds in rings can show substantial deviations from a linear geometry due to high ring strain (\textit{e.g.,} cyclooctyne, the smallest stable, cyclic hydrocarbon accommodating a triple bond, deviates by $\Delta \gamma =$ \SI{17}{\degree} from a linear geometry~\cite{bach2009ring}). For reference, we performed the same study on all molecules from the PubChemQC dataset~\cite{nakata2017pubchemqc} which include at least one non-ring triple bond (Figure~\ref{fig:linear_triple_bonds}), skipping molecules with could not be read with RDKit. From visual inspection of the distribution of triple bond angle deviations and manual inspection of example structures at different deviation levels, we decided to discard structures with a triple bond angle deviation of $\Delta \gamma >$ \SI{10}{\degree}. $1,147$ ($0.06$\%) conformations failed this test. Structures for which GFN2-xTB~\cite{grimme2017robust, bannwarth2019gfn2, grimme2019exploration, bannwarth2020extended} still indicated significant negative wavenumbers after 100 iterations ($17,502$ conformations, $0.88$\%) are denoted in the \texttt{summary.csv} file.

\noindent

\subsection{Deviation of aromatic rings from planar geometry}
\label{sec:planar_rings}
\begin{figure}[H]
\centering
\includegraphics[width=\textwidth]{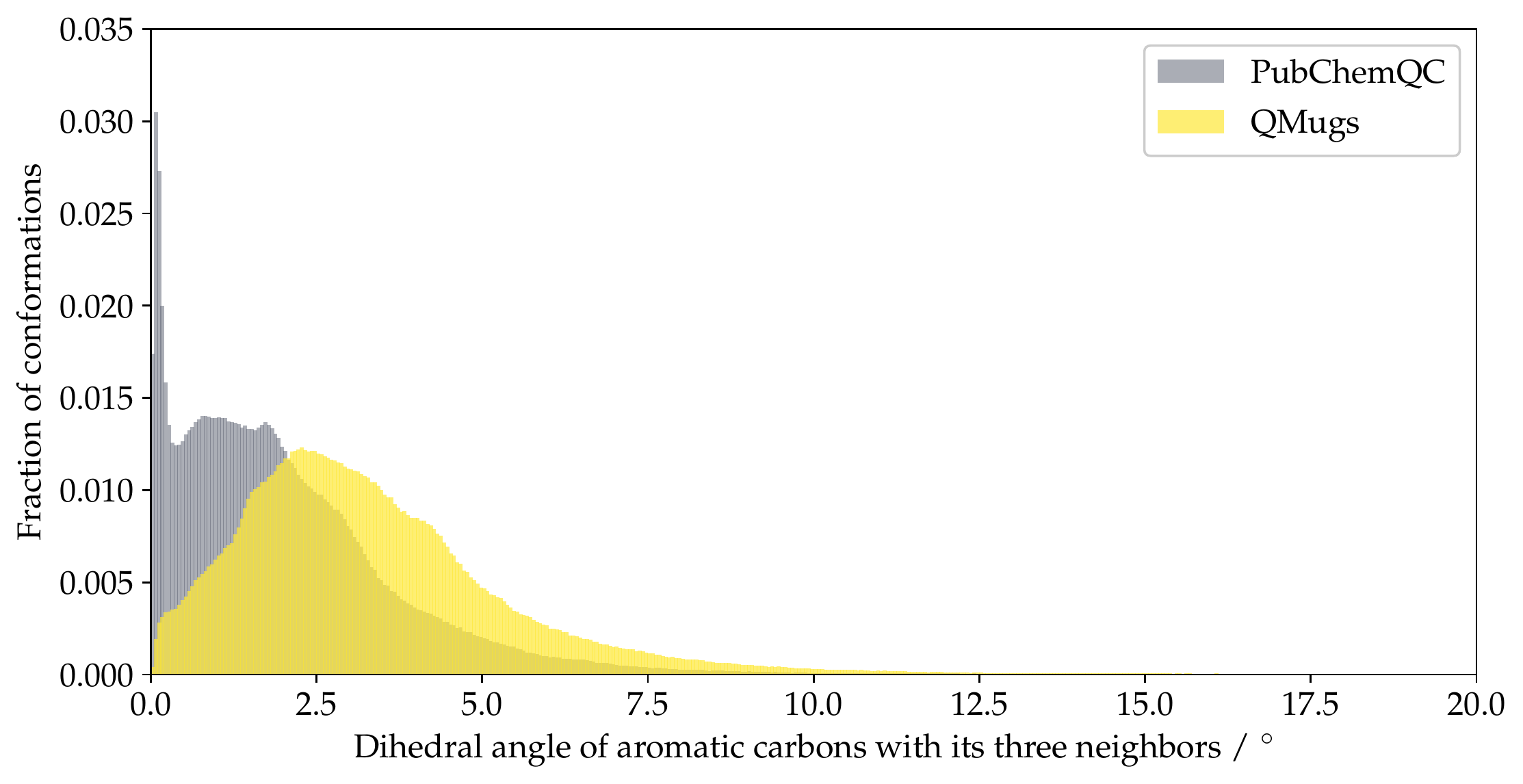}
\caption{Distribution of dihedral angle around aromatic carbons (histogram bin size \SI{0.05}{\degree}). Molecules with aromatic carbons from PubchemQC ($2,391,589$ conformations) and QMugs ($1,950,929$ conformations) show a deviation of $1.70$~$\pm$~\SI{1.85}{\degree} (median $\pm$ $1$ standard deviation) and $2.99$~$\pm$~\SI{2.20}{\degree}, respectively. $1050$ ($0.04$\%) molecules in the PubChemQC dataset and $564$ ($0.03$\%) molecules in the QMugs dataset with deviations greater than \SI{20}{\degree} are not shown.} 
\label{fig:carbon_C_angle}
\end{figure}

We furthermore investigated the planarity of carbon-containing aromatic rings. To do so, we assessed the dihedral angle between the two planes spanned by each aromatic carbon atom and its three neighbors. Angles greater than $90$\textdegree ~were corrected to 180\textdegree-<angle> to remove directional dependency. Calculations were performed for all order permutations of the aromatic carbon atom and its three neighbors. The largest dihedral angle (and hence the largest deviation from a perfectly planar aromatic ring) was recorded per conformation. We performed the same study on the molecules from the PubChemQC dataset~\cite{nakata2017pubchemqc} described in Section~\ref{sec:bond_length_check} (Figure~\ref{fig:carbon_C_angle}). From visual inspection of the distribution of aromatic carbon dihedral angle deviations
and manual inspection of example structures at different deviation levels, we decided to discard structures with an aromatic carbon dihedral angle deviation of \SI{15}{\degree} or more. $2,769$ ($0.14$\%) conformations failed this test.

\noindent

\newpage
\section{Independent terms of the Schrödinger equation}
\begin{figure}[H]
\centering
\includegraphics[width=\linewidth]{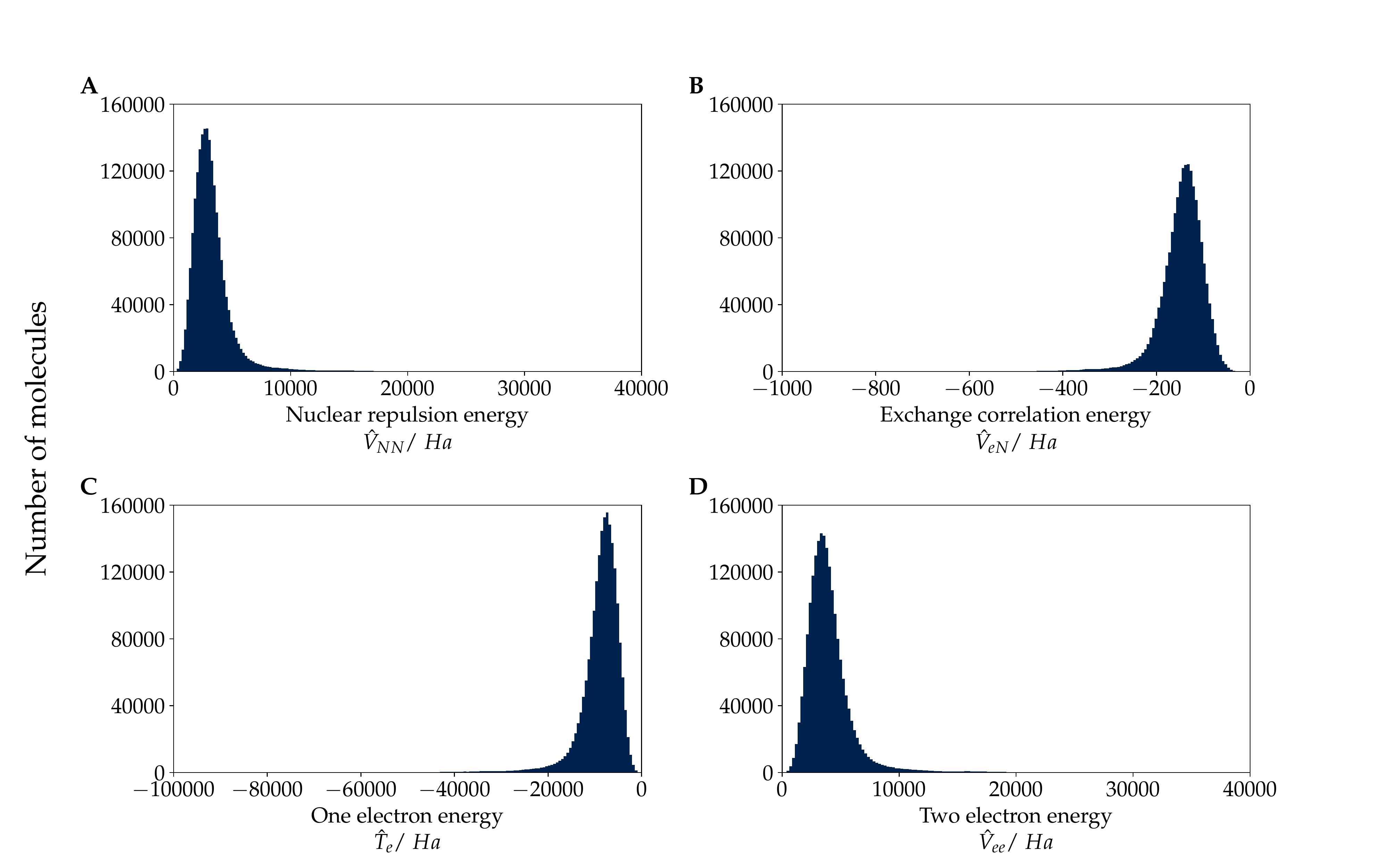}
\caption{Terms of the molecular hamiltonian $\hat{H}$ calculated on the $\omega$B97X-D/def2-SVP level-of-theory fo moelcules in QMugs. (\textbf{A}) Nuclear repulsion energy $\hat{V}_{NN}$ in $E_{H}$. (\textbf{B}) Exchange correlation energy $\hat{V}_{eN}$ in $E_{H}$. (\textbf{C}) One electron energy $\hat{T}_{e}$ in $E_{H}$. (\textbf{D}) Two electron energy $\hat{V}_{ee}$ in $E_{H}$. 
}
\label{fig:qm_se}
\end{figure}

The Born Oppenheimer approximation \cite{born1927quantentheorie} defines the molecular energy of the electronic Schrödinger equation as a sum of four independent terms, namely (i) nuclear repulsion energy $\hat{V}_{NN}$, (ii) exchange correlation energy $\hat{V}_{eN}$, (iii) kinetic electron energy also known as one electron energy $\hat{T}_{e}$, and (iv) electron repulsion energy also known as two electron energy $\hat{V}_{ee}$. The distribution of the four terms in QMugs are visualized in Figue \ref{fig:qm_se}.

\newpage
\section{Thermodynamic properties}
\begin{figure}[H]
\centering
\includegraphics[width=\linewidth]{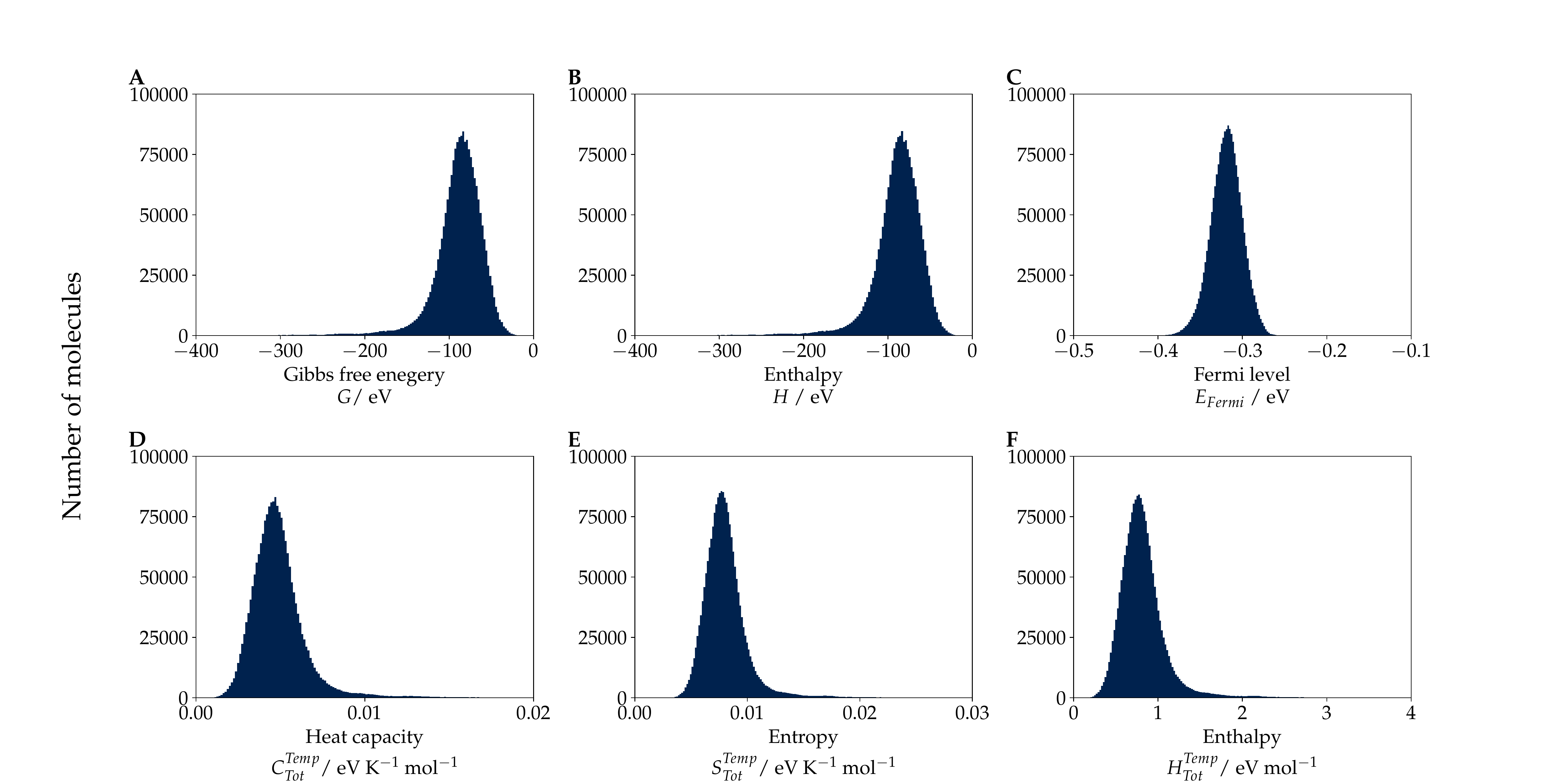}
\caption{Thermodynamic property distribution for molecules in QMugs calculated on the GFN2-xTB level-of-theory. (\textbf{A}) Gibbs free energy $G$ in eV. (\textbf{B}) Enthalpy $H$ in eV. (\textbf{C}) Fermi level $E_{Fermi}$ in eV. (\textbf{D}) Heat capacity $C_{Tot}^{Temp}$/ eV K$^{-1}$ mol$^{-1}$.  (\textbf{E}) Entropy $S_{Tot}^{Temp}$ in eV K$^{-1}$ mol$^{-1}$. (\textbf{F}) Partition function enthalpy $H_{Tot}^{Temp}$/ eV mol$^{-1}$.
}
\label{fig:theromo}
\end{figure}

Thermodynamic properties have been calculated for molecules in QMugs on the GFN2-xTB level-of-theory. Properties include total Gibbs free enegery $G$, total enthalpy $H$,  Fermi level $E_{Fermi}$, total heat capacity $C_{Tot}^{Temp}$, temperature dependent entropy $S_{Tot}^{Temp}$ and enthalpy $H_{Tot}^{Temp}$. $C_{Tot}^{Temp}$, $S_{Tot}^{Temp}$ and $H_{Tot}^{Temp}$ correspond to the sum of their individual rotational, transnational and vibrational terms, which can all be found in the SDF. Their distribution is shown in Figure \ref{fig:theromo}. 

\section{Additional figures and tables}

\begin{table}[ht]
\caption{Atom-type-specific atomic partial charge comparisons for the two levels of theory (GFN2-xTB, $\omega$B97X-D/def2-SVP) for the QMugs database. Abbreviations: RMSE, root mean squared error; PCC, Pearson's correlation coefficient.}
\centering
\begin{tabular}{@{}llll@{}}
\toprule
\textbf{Atom type} & \textbf{Occurrence} & \textbf{RMSE}          & \textbf{PCC} \\ \midrule
Hydrogen           & 49.0M  & $1.00 \times 10^{-3}$   & 0.913        \\
Carbon         & 45.1M  & 0.0130                   & 0.575          \\
Nitrogen         & 7.28M  & 0.0264                  & 0.124        \\
Oxygen           & 6.08M  & 0.0188                  & 0.274        \\
Fluorine         & 1.14M  & $3.21 \times 10^{-4}$   & 0.868        \\
Sulfur           & 729k  & 0.043                   & 0.991        \\
Chlorine         & 513k  & $6.39 \times 10^{-3}$   & 0.862        \\
Bromine          & 82.5k & 0.0153                   & 0.831        \\
Phosphorus       & 32.5k  & 0.139                   & 0.872        \\
Iodine           & 13.2k  & 0.0353                   & 0.808        \\ \bottomrule
\end{tabular}%
\end{table}

\begin{table}[ht]
\caption{Wiberg bond order comparisons  for the two levels of theory (GFN2-xTB, $\omega$B97X-D/def2-SVP) and the 15 most frequent pair-wise atomic covalent bonds in QMugs. Abbreviations: RMSE, root mean squared error; PCC, Pearson's correlation coefficient.}
\centering
\begin{tabular}{@{}llll@{}}
\toprule
\textbf{Bond type} & \textbf{Occurrence} & \textbf{RMSE}     & \textbf{PCC} \\ \midrule
Carbon-Hydrogen      & 44.8M  & $1.30 \times 10^{-3}$    & 0.787         \\
Carbon-Carbon        & 40.6M  & $6.83\times 10^{-4}$ & 0.998        \\
Carbon-Nitrogen      & 14.4M  & 0.0137             & 0.995        \\
Nitrogen-Hydrogen    & 3.20M  & $2.86 \times 10^{-3}$    & 0.860        \\
Carbon-Fluorine      & 1.14M  & 0.108             & 0.153        \\
Carbon-Sulfur        & 1.13M  & 0.0223             & 0.977        \\
Oxygen-Hydrogen      & 985k  & 0.0520             & 0.904        \\
Carbon-Oxygen        & 683k  & 0.0972             & 0.998        \\
Sulfur-Oxygen        & 618k  & 0.0853             & 0.941         \\ 
Nitrogen-Nitrogen    & 538k  & 0.0278             & 0.997         \\
Carbon-Chlorine      & 513k  & 0.0684             & 0.846         \\
Nitrogen-Sulfur      & 249k  & 0.0139             & 0.964        \\
Nitrogen-Oxygen      & 234k  & 0.110             & 0.997        \\
Phosphorus-Oxygen    & 107k  & 0.0112             & 0.980         \\
Carbon-Bromine       & 82.4k  & 0.0384             & 0.840        \\\bottomrule
\end{tabular}%
\end{table}

\begingroup
\renewcommand{\chapter}[2]{}
\newpage
\section*{References}
\bibliography{references_si}
\endgroup
\vspace{3cm}